\documentclass[twocolumn, apjl, twocolappendix, numberedappendix]{openjournal}
\usepackage{booktabs}
\usepackage{amsmath}
\usepackage{float}

\usepackage{times} 

\usepackage{graphicx}
\usepackage{amssymb}
\usepackage[dvipsnames]{xcolor}
\usepackage[backref, pagebackref=false, hyperindex=true, breaklinks=true, colorlinks=true, urlcolor=magenta, linkcolor=NavyBlue, citecolor=cyan, citebordercolor={0 1 0}, pagecolor=red, bookmarks=true, filecolor=blue,bookmarksopen=true, plainpages=false, pdfpagemode=UseThumbs]{hyperref}
\hypersetup{
colorlinks=true,
linkcolor=blue,
filecolor=blue,
urlcolor=blue,
citecolor=blue,
}
\usepackage{orcidlink}
\usepackage{multirow}
\usepackage{tabularx}

\newcommand{\yuti}{\texttt{Yuti}\,}
\newcommand{\bez}{B\'ezier\,}

\usepackage[dvipsnames]{xcolor}

\shortauthors{Bhowmick and Kumaran}

\begin{document}

\title{\textit{Beyond Spherical geometry}: Unraveling complex features of objects orbiting around stars from its transit light curve using deep learning}

\author{Ushasi Bhowmick$^1$ \orcidlink{0009-0007-4978-6445}}
\author{Shivam Kumaran$^1$ \orcidlink{0000-0002-9298-833X}} \vspace{1pt}
\affiliation{$^1$Space Sciences Division, Space Applications Centre, ISRO, Ahmedabad, Gujarat, India, 380015}
\email{ushasibhowmick@gmail.com}


\begin{abstract}

Characterizing the geometry of an object orbiting around a star from its transit light curve is a powerful tool to uncover various complex phenomena. This problem is inherently ill-posed, since similar or identical light curves can be produced by multiple different shapes. In this study, we investigate the extent to which the features of a shape can be embedded in a transit light curve. We generate a library of two-dimensional random shapes and simulate their transit light curves with light curve simulator, \yuti. Each shape is decomposed into a series of elliptical components expressed in the form of Fourier coefficients that adds increasingly diminishing perturbations to an ideal ellipse. We train deep neural networks to predict these Fourier coefficients directly from simulated light curves. Our results demonstrate that the neural network can successfully reconstruct the low-order ellipses, which describe overall shape, orientation and large-scale perturbations. For higher order ellipses the scale is successfully determined but the inference of eccentricity and orientation is limited, demonstrating the extent of shape information in the light curve. We explore the impact of non-convex shape features in reconstruction, and show its dependence on shape orientation. The level of reconstruction achieved by the neural network underscores the utility of using light curves as a means to extract geometric information from transiting systems.

\end{abstract}

\section{Introduction} \label{sec:intro}

The launch of exoplanet survey satellites such as Kepler and TESS has led to the discovery of more than 6000+ exoplanets from a set of more than 15000 project candidates. With increasing photometric precision of space-based telescopes, new possibilities have opened up for the characterization of diverse phenomena with the help of photometric transit light curves. 

The light curve of the system WASP 103b shows evidence of tidal distortion \citep{wasp103b_barros_2022}. The transits of $\beta$-Pictoris b show asymmetries corresponding to debris disk and exo-comets \citep{exocomet_etang_2022, exocomet_rene_2024}. \cite{eclipseInv_Williams_2006} theorized the use of eclipse light curves to obtain longitudinal flux variations for exoplanets. In the era of transmission spectroscopy of exoplanets, light curve anomalies are being used to infer asymmetric limb features \citep{limbassym_Paris_2016, limb_murphy_2024, limbwasp_murphy_2024, limbwasp_chen_2025}. \cite{LCandSpec_Yip_2020} and \cite{ PanRetrieve_Changeat_2024} demonstrate the utility of combining time domain along with spectral domain in atmospheric retrievals. This shows that anomalies in transit light curves can give us information about the geometric distortions in the transiting systems. 
 
The use of light curve inversion as a technique to reconstruct aspects of the shape geometry was first conceptualized for asteroid light curves \citep{ast_ostro_1988, ast_cellino_1989, kaasalainenI_2001, kaasalainenII_2001}. Asteroids are considered as 3D complex surfaces, and the light reflected from its surface is monitored over a given time period. The variations in the projected geometry over time gives rise to changes in flux. \cite{ast_ostro_1988, kaasalainenII_2001} demonstrate the detectability of convex hulls of asteroid shape. \cite{astell_muinonen_2015, astBayes_muinonen_2020, astBayes_muinonen_2022} establish an inversion model based on ellipsoid assumption and utilize it to perform retrievals of asteroid parameters from observed light curves. These methods have extended to the identification of near-Earth objects and space debris \citep{lcinvssa_bradley_2014, spdeb_jiang_2023, spdeb_galeano_2025}. Machine learning models are used extensively for optimizing inversion routines \citep{furfaro_2019, debris_james_2021, tang_2025}. \cite{tang_2025} uses deep neural networks to identify regions of non-convexity in reconstructed shapes. 

The technique has been extended for identification and characterization of starspots \citep{starspot_harmon_2000, latdist_santos_2017, starspotbi_luo_2019, spotrotdeg_luger_2021}. These works highlight the ill-posed nature of the inversion problem. In case of starspots, it manifests as a degeneracy between the number and location vs. the size of the starspots, especially when considering rotational systems \citep{spotrotdeg_luger_2021}. \cite{starspotbi_luo_2019} mentions the use of surface brightness ratio as priors to constrain the starspot reconstruction algorithms. These algorithms have been successfully used to constrain starspot properties for various systems, such as KIC 5110407 \citep{starspot_rottenbacher_2013}, Kepler-17b \citep{kep17_Luo_2021} etc. 

Recent studies on light curve inversion shows that the technique can be applied to transiting systems. \cite{oblatedeg_barnes_2003} simulates transits for oblate planets and demonstrate degeneracies with transit parameters. \cite{oblate_carter_2010} produce upper-bounds on oblateness of HD189733b. Pixel grid based approximation of transiting objects and its pixel-to-pixel shape reconstruction from light curves is also explored. \citet{eightbittransit_sandford_2019} approximates transit shapes using pixel-grid approach and demonstrates ambiguities such as flip, arch and stretch. \citet{exoringInv_arkhypov_2021} demonstrates inversion on y-symmetric shapes such as exorings. \citet{CODInversion_nachmani_2022} implements optimization techniques to reconstruct convex 2-D silhouettes.

\begin{figure*}[t!]
    \centering
    \includegraphics[width=1\linewidth]{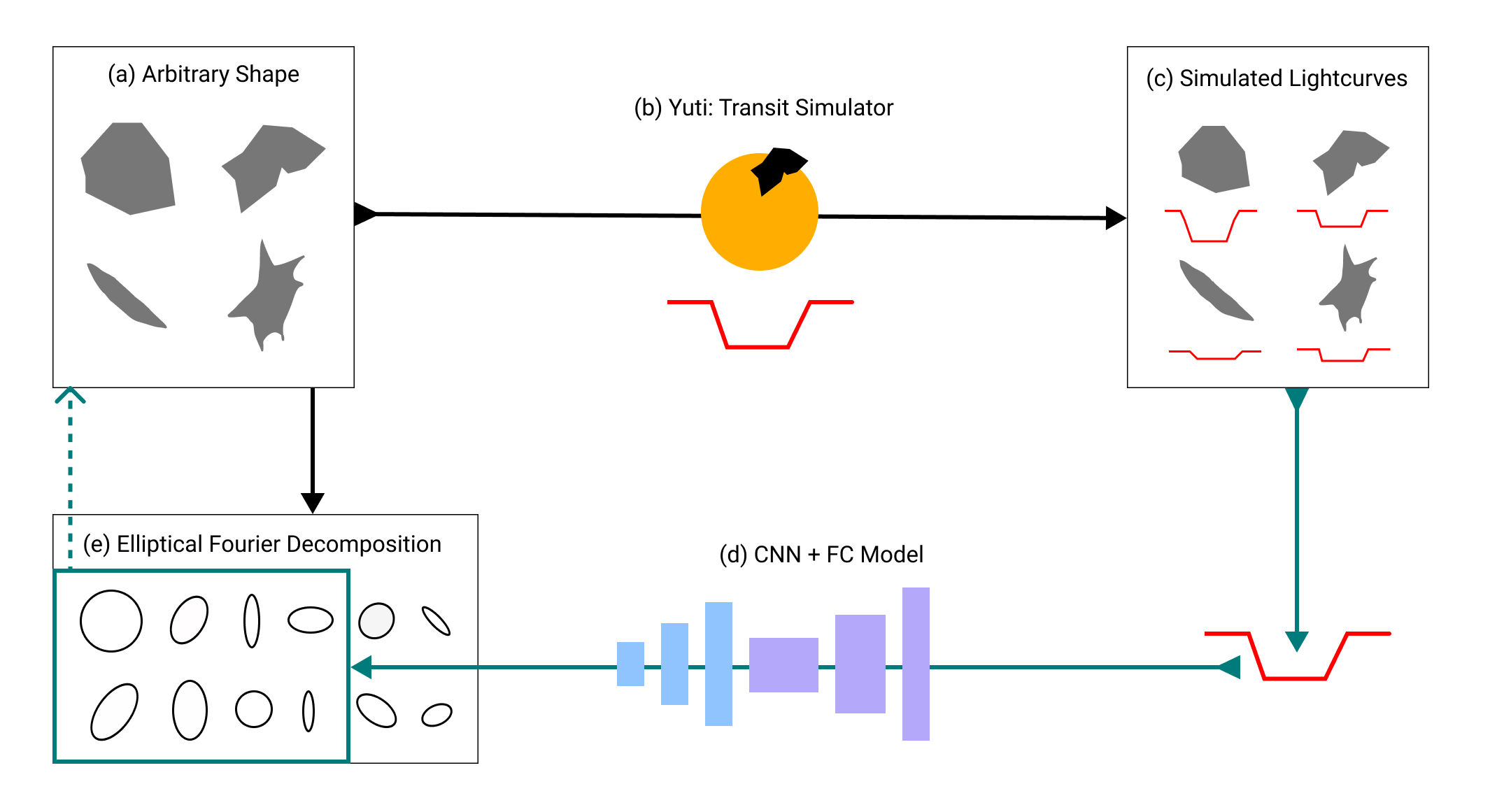}
    \caption{Methodology for inverting photometric transit light curves to recover 2-D shapes: (a) generate a comprehensive library of 2D shapes. (b) : using \yuti simulator ($\mathcal{Y}$) to generate corresponding transit light curves(c). (d) train neural-network to approximate the inverse function, $\mathcal{Y}^{-1}$. (e) Decomposition of 2D shapes into component ellipses and selection of subspace which can be predicted by the NN, effectively identifying invertible subspace (represented with dashed region)}
    \label{fig:methodology}
\end{figure*}

In this work, we investigate the extent to which features of a given shape are embedded in a transit light curve using machine learning. We generate a large number of arbitrary shapes, and use \yuti\footnote{\url{https://github.com/ushasi-bhowmick/Yuti}}  \citep{yuti_bhowmick_2024} to generate the transit light curve. Since inversion is an ill-posed problem, instead of a direct reconstruction of shape, we decompose the shapes into elliptical Fourier components. We train deep neural networks to learn the parameter space from the input transit light curve. Based on the ability of the neural network to learn the parameter values, we can make inferences about the feature space of shapes, and identify patterns that are detectable, and those that are ambiguous.

The paper is organized as follows. Section \ref{sec:2Dshapes} describes the methodology of generating arbitrary shapes and their elliptical Fourier decomposition. Section \ref{sec:MLTrain} describes the machine-learning model and the training strategies utilized. Section \ref{sec:first_order} shows the results of training for the first order Fourier coefficients and Section \ref{sec:higher_order} shows the results for higher order coefficients. Section \ref{sec:discuss} and \ref{sec:conclusion} presents the discussion and conclusions of the work.


\section{The Parameter Space of 2D Shapes} \label{sec:2Dshapes}

Photometric transit observations provide a one-dimensional imprint of the two-dimensional projection of a transiting object against the host star.  Consequently, the mapping from an arbitrary 2D shape $\mathcal{S}$ to the observed light curve $\mathcal{L}$ is intrinsically many-to-one as many different shapes can produce indistinguishable light curves. For example, shapes flipped along x-axis (the axis parallel to the direction of transit, passing through the center of the circular projection of a star) of the transit will result in an identical light curves. Some skewed features in the shape may not result in any change in the light curve from that of a shape without the protrusion. Such degeneracies pose a fundamental obstacle to an inversion algorithm that seeks to reconstruct the underlying shape solely from the observed light curve. To address this, we establish a methodology to identify the limits of recoverable information about the transiting shape embedded in the light curve. We generate a comprehensive library of 2D shapes and their corresponding light curves using the forward transit simulator \yuti, denoted here as $\mathcal{Y}$. By leveraging the universal function‑approximation ability of neural networks \citep{Hornik1989}, we train several models to learn the inverse mapping \(\mathcal{Y}^{-1}\) from a large library of shapes paired with their light curves.  Each network receives a light curve as input and extracts only those features that survive the many‑to‑one projection, thereby revealing the shape characteristics that are, in principle, recoverable.

\begin{figure*}
    \includegraphics[width=1\linewidth]{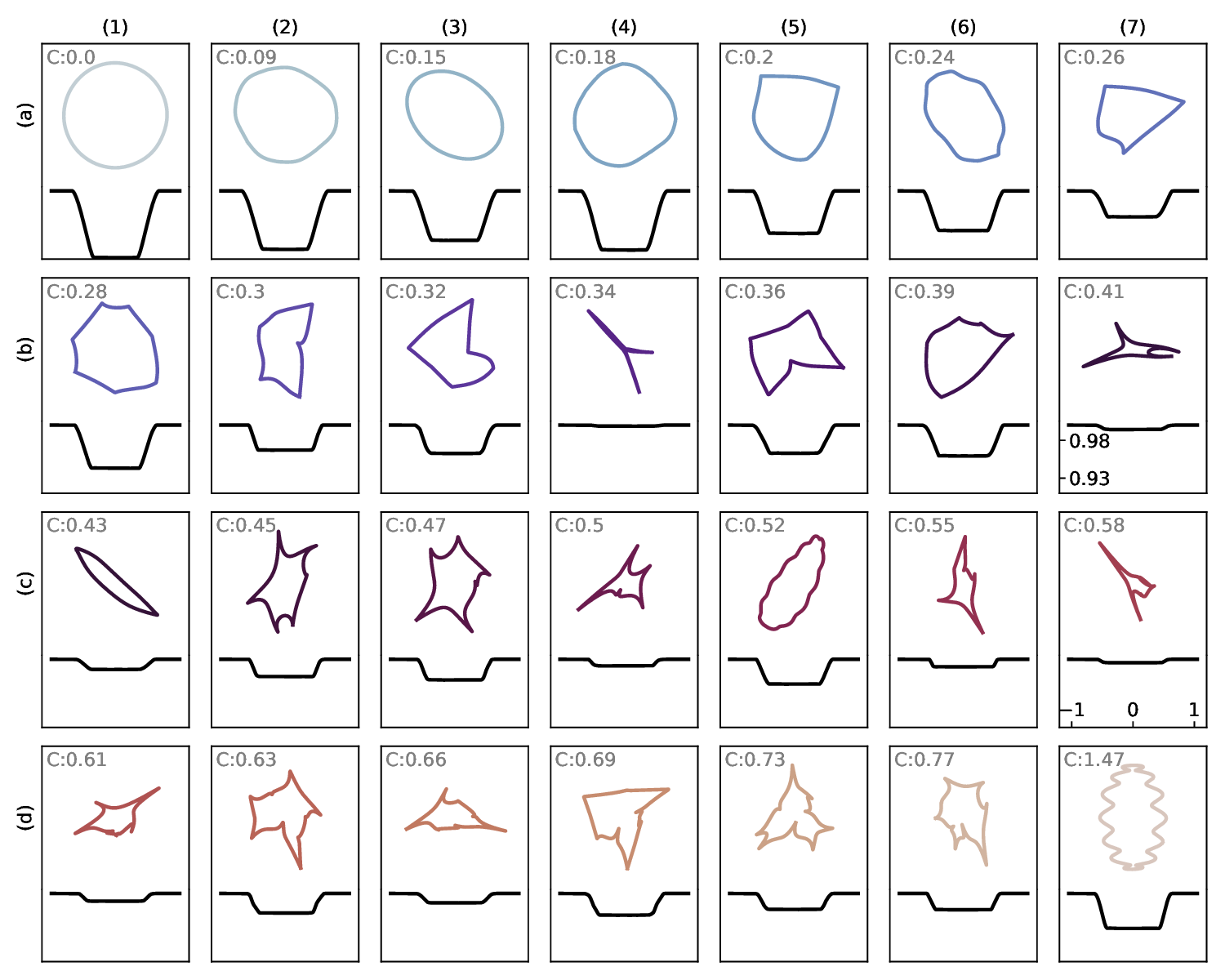}
    \caption{Examples from the training sample are shown. The top part of each panel shows the shapes, arranged in increasing order of complexity. The complexity values are mentioned in the top left corner. The bottom part of each panel shows the corresponding light curve generated using \yuti.  
    \label{fig:sh_c_eg}}
\end{figure*}

Fig. \ref{fig:methodology} explains the overall methodology as per the following steps:
\begin{itemize}
    \item Generate a library of arbitrary 2D shapes (panel-a).
    \item Obtain the transit light curve of these 2D shapes using \yuti simulator (panel-b $\rightarrow$ panel-c).
    \item Find low dimensional embedding of the input shapes (panel-e).
    \item Design and train ML models to predict the low dimensional shape embedding using the light curve given as input to the model (panel-d).
    \item Analyze the subspace in which the ML model is successfully able to map the inverse function of \yuti simulator, which is essentially the subspace in which the light curve is non-degenerate w.r.t. the 2D shapes (represented by shaded region in panel-e).
    \item Using the low-dimensional decomposition, identify the features which can be successfully obtained from a transit light curve of an unknown shape.
\end{itemize}

The following subsections describe the generation of the 2D shape library, and the strategy for creation of a low dimensional embedding using elliptical Fourier descriptors.

\subsection{2D shapes generation by \bez Curves}
  For supervised training of the machine learning model(s), we need to generate a large number of arbitrary random shapes. \bez curves are widely used to parameterize 2D profiles and generate random shapes for various applications \citep{lesion_oliviera_2020, laminar_viquearat_2020, dem_wang_2021}. 
  We use the algorithm developed by \cite{bezier_viquerat_2021}\footnote{\url{https://github.com/jviquerat/shapes}} where we define a set of control points and an associated \texttt{rad} and \texttt{edgy} parameter for each point. A cubic \bez curve is drawn between adjacent control points to create a 2D shape. We generate a large number of random 2D shapes by sampling uniform random values of (x, y) control points, and \texttt{rad, edgy} parameters.  Each generated shape is shifted and scaled to have centroid at $(0,0)$ and its outermost point on the unit circle. This is done, so that all shapes are at the same scale, and there are no effects of size of the shape in the transit light curve.

  This algorithm can lead to the generation of a large number of similar-looking shapes. For example, to generate an ideal circle, infinite points are needed. Therefore, a number of different sets of control points can lead to very circular shapes. Three control points with different locations can lead to triangular shapes of different orientation. In order to obtain a training sample with the appropriate diversity of shapes, we evaluate each \bez shape with a complexity metric. 

  \cite{complexity_chen_2005} define a complexity metric (referred to as $\mathcal{C}$ in this text) for a 2D shape based on a global distance entropy, local angle entropy, perceptual smoothness and a randomness factor. The transit light curve is influenced by the silhouette or the 2D projection of the 3D shape. A spherical exoplanet has a circular projection, which has $\mathcal{C}=0$. If an object's projection has uneven boundaries, like irregularly shaped asteroids, such 2D projections can have complexity in the range $0.2-0.4$. On the other hand, certain projections of artificial satellites may have complexity greater than $1$. Therefore, the complexity metric is a good representation for complex 2D projections. For each random shape in our dataset, we evaluate the corresponding complexity values. Fig. \ref{fig:sh_c_eg} shows samples from the set of generated shapes. The complexity value is also mentioned. The higher the complexity value, the more visually complex the shape appears. In order to capture the visual variation of generated shapes, we select a set of $20000$ shapes for our training sample, such that the training sample is uniformly distributed in the complexity space.

\subsection{Transit light curves}
  In the previous section, we generated a large number of random shapes for our training sample. We use \yuti to simulate the corresponding transit light curves. For the purpose of this study, we keep the transit parameters fixed. The scale of the shapes are fixed, i.e. $R_p/R_{st} = 0.3$ (size of the unit circle w.r.t stellar radius). Orbital Distance is taken $3 R_{st}$ and impact parameter is $0$. We assume that the object's projected area remains invariant during the transit, which implies that we do not consider rotating objects or objects undergoing transient deformations. In addition, we assume that the star has no limb darkening. These assumptions are made to simplify the inversion problem, and enable a more efficient exploration of the Fourier parameter space. We simulate each transit with $~5$ million Monte-Carlo Points (see Appendix \ref{apdx:mc-noise}). We sample an orbital phase of $2\pi/3$ by 300 points around the center of the star during the transit. The simulated transits are shown in the bottom part of all panels in Fig. \ref{fig:sh_c_eg}. 

\subsection{Shape decomposition: Low Dimensional Embedding}
  A direct reconstruction of shape from the transit light curve, will lead to challenges in identifying the degeneracies in the light curve. ML models inherently assume a functional relation between inputs and outputs; therefore, the model will predict one of the many possible shape outcomes. Therefore, we need to create a low-dimensional embedding of the shape parameter space, where each parameter quantifies a certain aspect of the shape. The Complexity metric($\mathcal{C}$) itself can be considered as a one-dimensional latent vector. We can train a machine-learning model to predict the complexity of a transiting shape from the light curve input. The Complexity value itself gives us a good idea of the anomalous nature of the transiting object. 

  A single value captures very limited information about an anomalous shape; therefore, we require a strategy to generate a parameter space with more dimensions. We explore a number of different strategies such as Principle Component Analysis (PCA) and Auto-encoders to create a low dimensional embedding. However, statistics-based strategies such as PCA are generated based on the training sample itself, and therefore is more likely to be biased by our initial strategy to generate these shapes. In addition, this poses limitations on the interpretability of the parameter itself. In order to obtain an interpretable and generalised embedding we use elliptical Fourier descriptors as defined by \cite{ellipdesc_kuhl_1982} to define the shape.

  Using elliptical Fourier descriptors (EFD) up to $N^{th}$ order, a 2D closed curve $(x, y)$ sampled with $M$ boundary points \(\{(x_i,y_i)\}_{i=0}^{M-1}\) is parameterized as follows using $t$ as sampling parameter:

  \begin{equation}
      x(t) = \sum_{n=1}^N c^{n}_1 cos( 2n\pi t ) + c^{n}_2 sin(2n\pi t)
      \label{eq:fr_x}
  \end{equation}

  \begin{equation}
      y(t) = \sum_{n=1}^N c^{n}_3 cos(2n\pi t) + c^{n}_4 sin(2n\pi t)
      \label{eq:fr_y}
  \end{equation}

  Where $c^{n}_1, c^{n}_2, c^{n}_3, c^{n}_4$ are Fourier coefficients for the $n^{th}$ order. The equations of $x(t)$ and $y(t)$ does not contain constant terms as all the shapes are shifted such that centroid of the shape is at the origin. A shape can be represented with sufficient accuracy for a finite $N$, nevertheless as $N\rightarrow \infty$, the Fourier representation becomes exact as the given shape. Because Fourier coefficients depend on the sampling of the shape, we sample each shape in the training set uniformly, with equal number of points. Each Fourier order defines an ellipse on the 2D plane, where the $n^{th}$ order ellipse is traversed $n$ times. Therefore, the Fourier coefficients for each order can be converted to equivalent parameters of a rotated ellipse to obtain the following description: 

  \begin{equation}
        \begin{bmatrix}
            x \\
            y
        \end{bmatrix} = \sum_{n=1}^N
        \begin{bmatrix}
            -sin\Theta_n & cos\Theta_n \\
            cos\Theta_n & sin\Theta_n \\
        \end{bmatrix} 
        \begin{bmatrix}
            a_n cos(2n\pi t + \phi_n) \\
            b_n sin(2n\pi t + \phi_n)
        \end{bmatrix}
  \end{equation}

  Where $a_n$ is the semi-major axis, $b_n$ is the semi-minor axis, $\Theta_n$ is the angle of orientation of semi-major axis, and $\phi_n$ defines the starting point on the ellipse for a given Fourier order $n$. All Fourier coefficients in this description have bounds. $a$ varies between $(0,1)$ and $b$ varies between $(-1, 1)$. For a given order, the sign of $b$ determines if the ellipse is traversed clockwise or anticlockwise. Flipping the sign of all orders of $b$ leads to the same shape, but traversed clockwise instead of counter-clockwise (or vice versa). This has no impact on the silhouette, therefore, we choose $b_1$ to take values between $(0,1)$, and rest of the orders to have values between $(-1,1)$. For an ellipse, $\Theta$ can take any value within a $2\pi$ interval, i.e. ($-\pi, \pi$). We choose an interval between $(-\pi/2, \pi/2)$ for $\Theta$. Angles outside this interval lead to formation of same ellipse, but with different starting point ($\phi$). Finally, $\phi$ can vary between $(-\pi, \pi)$. The conversion between the two descriptions is given in Appendix \ref{apdx:fourier-conversion}.

  \begin{figure}
     \centering
     \includegraphics[width=1\linewidth]{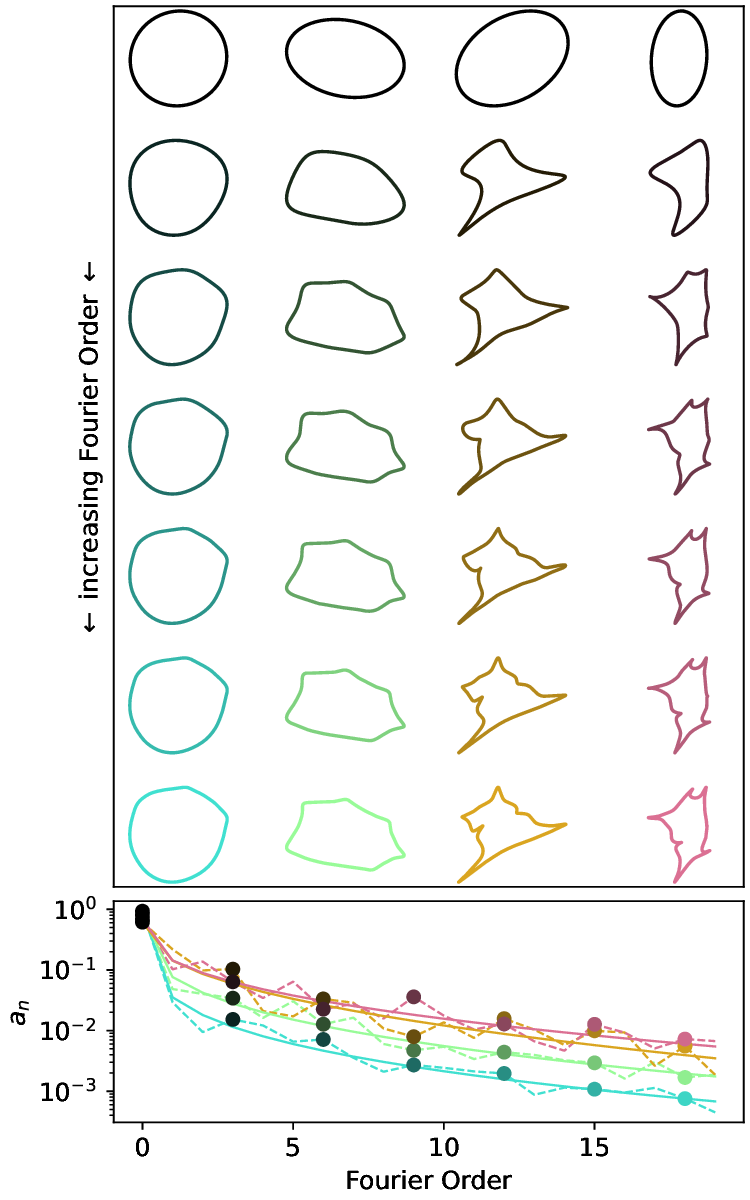}
     \caption{Illustrative example of Fourier decomposition for 4 selected shapes with increasing complexities.  The top panel shows the evolution of the shape with increasing number of Fourier Orders. The bottom panel shows the variation of $a_n$ for each shape. The dashed lines represent the calculated $a_n$ values in each order, and the solid line shows the fitted trend. The markers correspond to individual shapes in the top panel.}
     \label{fig:shape-shifter}
  \end{figure}

  \begin{figure}
    \includegraphics[width=1\linewidth]{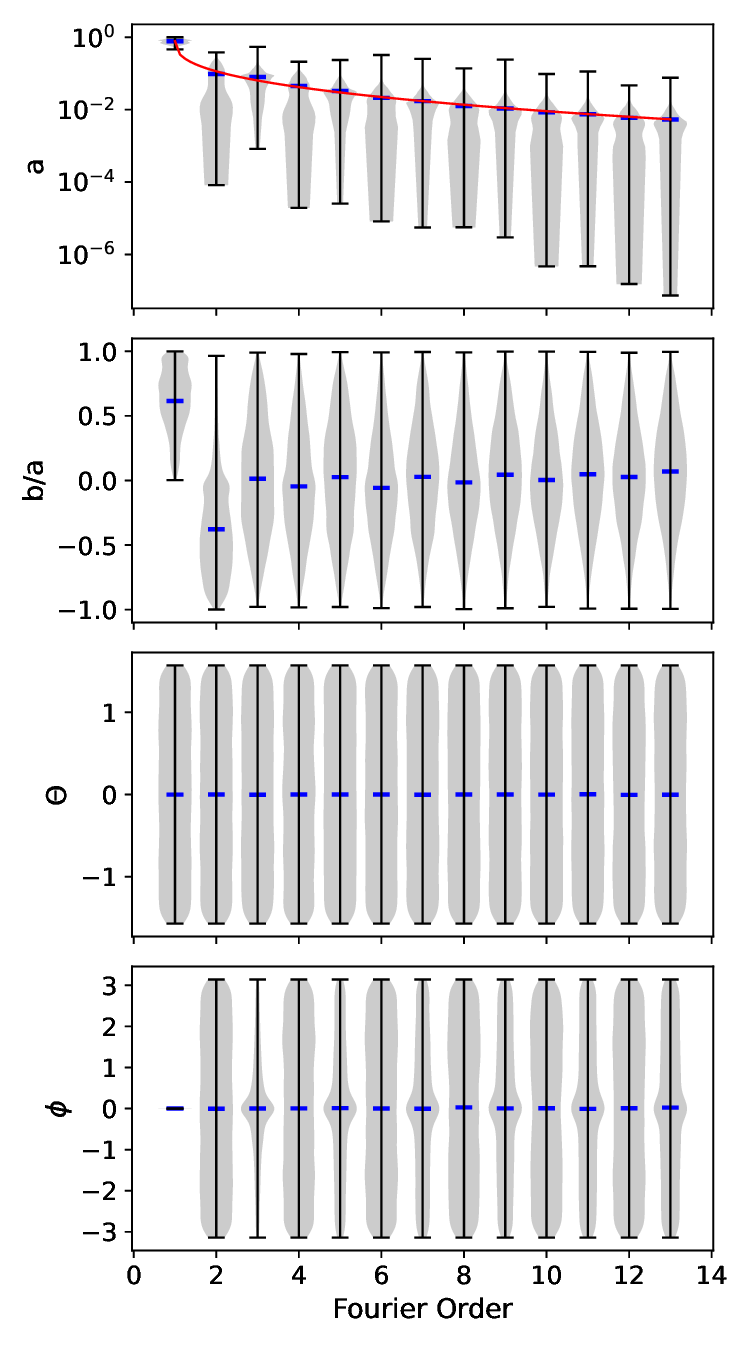}
    \caption{The distribution of parameters in the training sample for various Fourier orders is shown. The 1st panel shows the distribution of $a$ which follows a logarithmic trend. The trend followed by the mean of $a$ across various orders is shown in red. The $2^{nd}$ panel shows the distribution of $b/a$. The third panel shows the distribution of $\Theta$, and the fourth panel shows the distribution of $\phi$. The blue markers represent the mean values of individual distributions.
    \label{fig:par_dist}}
  \end{figure}
  
  We calculate the Fourier coefficients $(a_n, b_n, \Theta_n, \phi_n)$ $\forall n \in [0, N]$ for each shape in our training sample. The Fourier decomposition of 4 sample shapes are shown in Fig. \ref{fig:shape-shifter}. With increasing number of Fourier orders, the reconstructed shape comes closer to the original shape. The $a_n$ values for the shapes across Fourier orders is shown in the bottom panel. The $a_n$ parameter is observed to follow a logarithmic decline across Fourier orders. The decline is steeper for a less complex shape, and more gradual for a highly complex shape. This trend is represented empirically by the following equation:
  
  \begin{equation}
      log(a_n) = \Lambda_a(n-1)^{\gamma_a} + log(a_1)\label{eq:trend}
  \end{equation}

  Where $\Lambda_a, \gamma_a$ are parameters that can be used to estimate  the higher order $a_n$ for a shape. The bottom panel of Fig. \ref{fig:shape-shifter} show the calculated $a_n$ values and the corresponding trend line of the shape.

  The distribution of the parameters $a, b, \Theta, \phi$ for $13$ Fourier orders is shown in Fig. \ref{fig:par_dist}. Similar to the individual $a_n$, it's distribution (mean value) also follows a logarithmic trend in Fourier orders with $log(a_1) = -0.255$, $\Lambda_a = -1.914$ and $\gamma_a = 0.383$ (from eq. \ref{eq:trend}). Even order parameters ($a_2, a_4, a_6$...) show larger variation in values than the odd orders. Since the semi-minor axis $b$ also incorporates the trends followed by $a$, we use the parameter $b/a$ for this analysis. This gets rid of the logarithmic trend, as seen in the distribution of $b/a$ in the second panel of Fig. \ref{fig:par_dist}. $b_1$ values ranges from 0 to 1, while higher orders can be both positive and negative, spanning -1 to 1.
  
  The $\Theta$ parameter is consistent between ($-\pi/2, \pi/2$) across all Fourier orders. The $1^{st}$ order in $\phi$ is $0$ for all shapes. This is by design, since the starting point of drawing the 2D shape is not relevant to the projected overlapping geometry, therefore, all shapes can be made $\phi$ invariant in the first order. For higher orders, $\phi$ varies between $(-\pi, \pi)$. We see that the variation in values is greater for even orders ($\phi_2, \phi_4$..) and lesser for odd orders ($\phi_3, \phi_5$...). 

  \begin{figure}
    \includegraphics[width=1\linewidth]{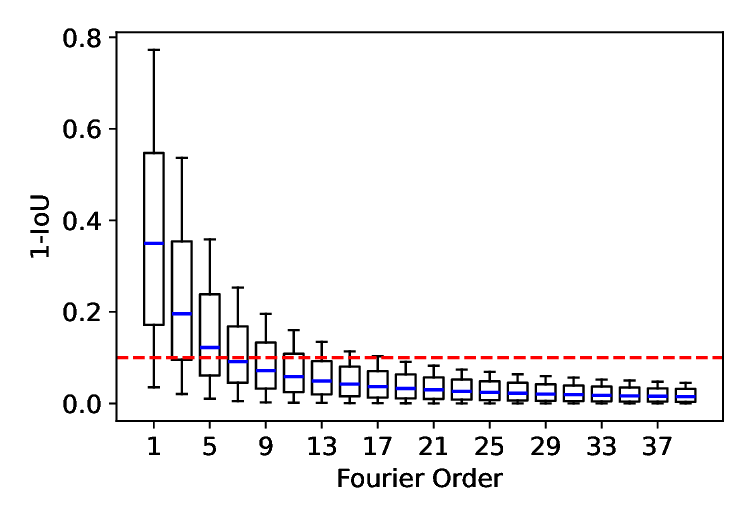}
    \caption{The distribution of reconstruction error for various Fourier orders. For each order, reconstruction error is calculated between the original shape and the shape obtained by truncating the Fourier series at the given order. The blue markers represent the median value. The boxes represent the 25 percentile (lower) and 75 percentile (higher). The whiskers represent the 5 percentile(lower) and 95 percentile (upper). The dashed red line represents $1-IoU=0.1$.
    \label{fig:trunc_dist}}
  \end{figure}

  In order to quantify the similarity between two shapes, we calculate the Intersection over Union ($IoU$) of uniformly rasterized shapes. We take $1-IoU$ as reconstruction error. An ideal Fourier series consists of infinite orders. As we incorporate increasing number of Fourier orders, the reconstructed shapes become more similar to the original shape and reconstruction loss decreases. Fig. \ref{fig:trunc_dist} shows the decrease in reconstruction loss across our training sample for increasing Fourier orders. For each order, the reconstructed shape is calculated by truncating the Fourier series at the given order for each shape in the training sample. We see that the median reconstruction error decreases rapidly as we go to higher order coefficients. By $N=19$, $95\%$ of shapes have $IoU$ better than $0.9$. Therefore we choose to truncate the Fourier series at order $N=20$ for all our shapes.
    
  In addition to the \bez generated 2D shapes, we create additional 2D shapes by random uniform sampling of EFDs. The EFD sampling results in self-intersecting (loopy) curves, which to not correspond to realistic projection of transiting objects. After discarding such loopy shapes, we add 10000 other 2D shapes to the training sample. This shape-set is also kept uniform in complexity, in line with our previous \bez generation. Finally we get a total of $30000$ shape and corresponding \yuti generated light curves as the training sample.
    
  A transiting shape and its flip along the x-axis of transit have identical light curves. Since flip degeneracy is expected to manifest in training, it is necessary to understand the impact of this in the Fourier decomposition. In equation \ref{eq:fr_x}, \ref{eq:fr_y}, replacing $y$ with $-y$ gives us the x-axis-flipped shape (assuming impact parameter of $0$). In the elliptic description this amounts to $\Theta_n \rightarrow -\Theta_n$ and $\phi_n \rightarrow -\phi_n  \forall n$. This means that the training of $\Theta, \phi$ for individual Fourier orders is likely to be affected by flip degeneracy. 


\section{ML Model Training} \label{sec:MLTrain}

\begin{figure}
    \centering
    \includegraphics[width=1\linewidth]{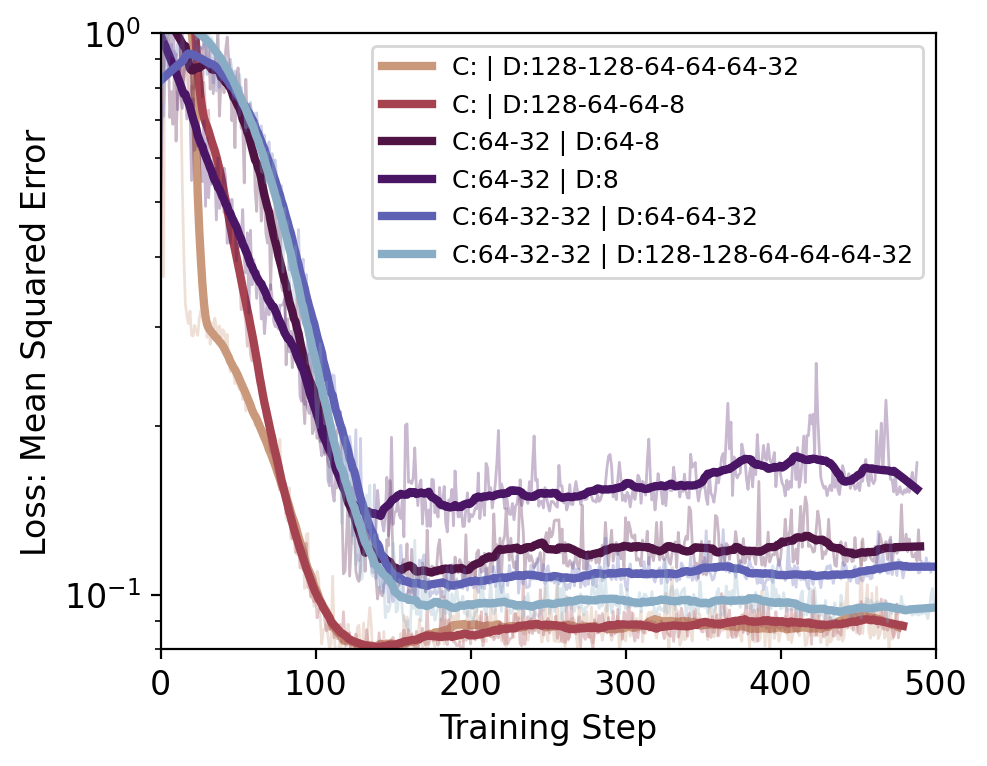}
    \caption{Mean squared error on validation dataset for various model architectures for parameter $a_2$. The legends show the associated model: Convolution Layers, followed by dense layers (denoted as C and D respectively)\\}
    \label{fig:training-loss-curve}
\end{figure}

\begin{table}
    \centering
    \begin{tabular}{lllrr}
    \toprule
    \textbf{Name} &          \textbf{ConvL} &                   \textbf{DenseL} &  \textbf{r} &  \textbf{m} \\
    (1)  &          (2)    &                    (3)  &  (4)  & (5)\\
    \midrule
        $\mathcal{C}$ & & 128,128,64,64,64,32 & 0.73 & 0.80 \\
        $a_1$ & 64,32 & 64,64,32 & 0.96 & 1.01 \\
        $a_2$ &  & 128, 64, 64, 8 & 0.93 & 0.86 \\
        $a_3$ &  & 128, 64, 64, 8 & 0.85 & 0.94 \\
        $a_4$ &  & 128, 64, 64, 8 & 0.90 & 0.84 \\
        $a_5$ & 64, 32 & 64, 8 & 0.69 & 1.05 \\
        
        $b_1$ & 64, 64, 32, 32 & 64, 8 & 0.99 & 1.0 \\
        $b_2/a_2$ &  & 128,128,64,64,64,32 & 0.60 & 0.78 \\
        $b_3/a_3$ &  & 128,128,64,64,64,32 & 0.71 & 0.88 \\
        $b_4/a_4$ &  & 128,64,64, 8 & 0.54 & 0.66 \\
        $b_5/a_5$ &  & 128,64,64, 8 & 0.54 & 0.43 \\
        
        $\Gamma_a$ &  & 128,64,64,8 & 0.71 & 0.75 \\
        $\Lambda_a$ &  & 128,64,64, 8 & 0.92 & 1.00 \\
        
        $|\Theta_1|$ & 64, 64, 32, 32 & 64, 64, 32 & 0.95 & 0.98 \\
        $|\Theta_2|$ & 64, 32, 32 & 128, 128, 64, 64, 8 & 0.54 & 0.74 \\
        $|\Theta_3|$ & 32 & 128, 64, 64, 8 & 0.62 & 0.79 \\
        $|\Theta_4|$ &  & 128, 64, 64, 8 & 0.48 & 0.59 \\
        $|\Theta_5|$ &  & 128, 64, 64, 8 & 0.32 & 0.17 \\
        $|\phi_2|$ &  & 64, 64, 32 & 0.58 & 0.58 \\
        $|\phi_3|$ &  & 64, 64, 32 & 0.48 & 0.55 \\
        $|\phi_4|$ &  & 128,128,64,64,64,32 & 0.38 & 0.38 \\
        $|\phi_5|$ &  & 128,128,64,64,64,32 & 0.41 & 0.25 \\[4pt]

          & & \multicolumn{3}{r}{\textbf{Accuracy}} \\
        $s[\Theta_1]$ &  & 128,64,64, 8 & \multicolumn{2}{r}{0.90} \\
        $s[\Theta_2]$ &  & 128,128,64,64,64, 32 & \multicolumn{2}{r}{0.69} \\
        $s[\Theta_3]$ &  & 64, 64, 32 & \multicolumn{2}{r}{0.74} \\
        $s[\Theta_4]$ &  & 64, 64, 32 & \multicolumn{2}{r}{0.62} \\
        $s[\Theta_5]$ & 64, 32 & 8 & \multicolumn{2}{r}{0.58} \\
        $s[\phi_2]$ &  & 128, 64, 64, 8 & \multicolumn{2}{r}{0.70} \\
        $s[\phi_3]$ &  & 64, 64, 32 & \multicolumn{2}{r}{0.65} \\
        $s[\phi_4]$ &  & 64, 64, 32 & \multicolumn{2}{r}{0.61} \\
        $s[\phi_5]$ &  & 64, 64, 32 & \multicolumn{2}{r}{0.55} \\
    \bottomrule
    \end{tabular}
    \caption{Architecture of best neural Network for individual elliptical parameters up to $5^{th}$ order. Columns: (1) parameter name, (2) Number of filters in convolution layers, (3) Number of hidden nodes in the dense layers, (4) Pearson R coefficient between true and predicted parameter values, (5) slope of the fitted line between true and predicted values. The parameters names: $s[\Theta_n]$ and $s[\phi_n]$ represents the sign of $\Theta_n$ and $\phi_n$ respectively. Since the sign are binary values, `Accuracy' is taken as selection metric for best neural-network.}
    \label{tab:model-selection}
\end{table}

Using elliptic Fourier descriptors, the arbitrary 2D shapes are broken down into component ellipses. The shape is thus represented (and can be reasonably reconstructed) by taking top N elliptical parameters: $a_n$, $b_n$, $\Theta_n$, $\phi_n$. The light curves generated using the \yuti simulator (refer Fig. \ref{fig:methodology} (b)) are 1D arrays of 300 elements. With the ML model, we need to predict the elliptical parameters ($a_n$, $b_n$, $\Theta_n$, $\phi_n$) up to a sufficient order (see Fig. \ref{fig:trunc_dist}). This sufficiency in order may not be uniform across all the parameters, and is dependant on the scale of values. For example, the $a_n$'s show a logarithmically declining trend (Fig. \ref{fig:par_dist}). The Fourier coefficients influence the light curve very unevenly. The $a_n$ and $b_n$ terms set the overall size, eccentricity and orientation of effective ellipse, while $\Theta_n$ and $\phi_n$ encode finer angular details. The contribution of each coefficient decreases with increasing order. In addition, this decline across order is not uniform across the coefficients. Because of this disparity, a single neural network (NN) would need to be supplied with corresponding weights for every parameter. These weights must be assigned before training and would require extensive fine-tuning of output scaling or loss terms. Training an independent network for each coefficient sidesteps this problem. Instead of representing the relative influence of individual parameter, the entire NN architecture can be optimized for the influence of a single coefficient on the light curve. Although this increases the number of hyper-parameter tuning searches, the limited set of useful orders keeps the total number of networks small while still capturing the shape information embedded in the light curve.
The general architectures of the neural network models have combination of Convolution layers followed by Dense layers (see Fig. \ref{fig:methodology} (d)). The hyper-parameter tuning space for the layer architectures is:

\begin{itemize}
    \item Convolution layers layers: [\,], [32], [64,32], [64,32,32], [64,64,32,32]
    \item Dense Layers : [8] , [64,8], [64,64,32], [128,64,64,8], [128,128,64,64,64,32]
\end{itemize}
    
Here `[\,]' indicates no convolution layer. For a given parameter, we construct all possible network architectures from the above list of Dense and Convolution layers. The input light curve is normalized such that the values are between 0 and 1. The output elliptical parameters are also normalized between 0 and 1. At the output activation layer, we use \textsc{sigmoid} activation function. For one selected architecture, the model is trained using Adam optimizer with a learning rate of 0.0005.  The light curve and the elliptical parameters are split into 0.8-0.2 fractions, 0.8 for training set and 0.2 held-out as test set. The 0.8 training set is split into another 0.8:0.2 training:validation set. This validation set is used to compute validation loss at the end of each epoch. To avoid overfitting, we use early-stopping, where the training is halted as soon as the validation loss ceases to improve and begins to rise relative to the training loss. To avoid halt due to short fluctuations in validation loss, we keep a patience of 10 epochs in early stopping. To further mitigate overfitting, we include a regularization via a dropout layer (10\% drop rate) after the CNN layer and immediately before the dense layers (after batch normalization layer in case of dense-only network).

\begin{figure}
    \includegraphics[width=1\linewidth]{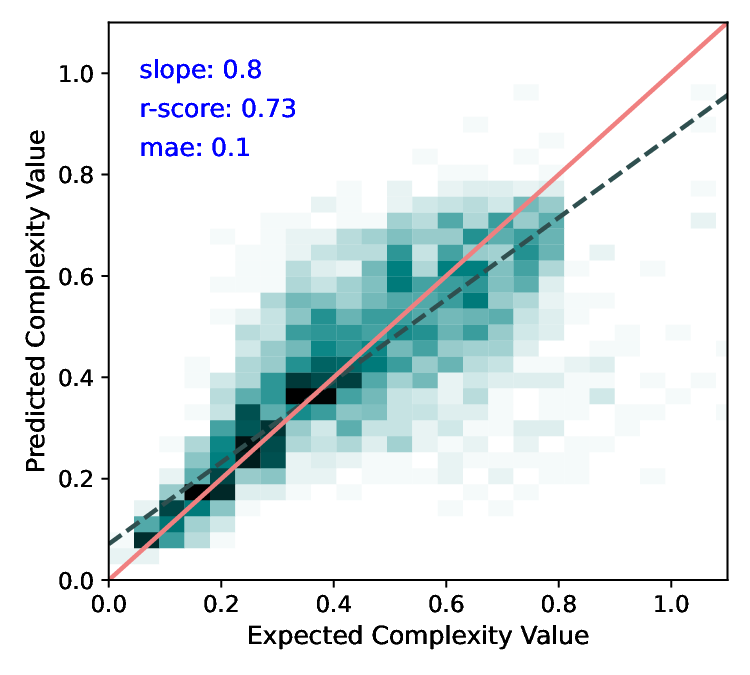}
    \caption{ Predicted value of complexity vs expected value of complexity from the trained ML model. The solid pink line shows the 1-1 relation, whereas the dashed line shows the slope of the best fit line to the test sample. $r$ represents the Pearson coefficient, and mae corresponds to the mean absolute error.
    \label{fig:complex_train}}
\end{figure}

The best model for a given parameter is identified using the combination of Pearson correlation coefficient (referred to as r-score or $r$ in this work) and the slope of best fit line ($m$). These matrices are computed on the held-out test set which was not used while training and validation. For obtaining the slope, we use Deming regression \citep{Deming1943} over the true values in the test data and the model predicted values. The r-score measures the linearity between two given datasets. Greater values of $r$ imply a strong correlation between expected and predicted parameters. However, it does not determine the the accuracy of the predicted value. Therefore, we also analyze the slope. Values of slope ($m$) closer to $1$ imply that the predicted parameter matches the expected parameter closely. As an example, the training loss curves obtained for different models during hyperparameter tuning of $a_2$ are shown in Fig. \ref{fig:training-loss-curve}. We see that both models D:128-128-64-64-64-32, and model D:128-64-64-8 show similar performance in training. This indicates that a more complex network is not needed to improve training. Table \ref{tab:model-selection} lists out the best models for each parameter obtained after hyperparameter tuning.

The following sections present the detailed analysis for individual Fourier parameters. 
    

\section{Predictions of Extent and Orientation} \label{sec:first_order}

\begin{figure*}[ht]
    \includegraphics[width=1\linewidth]{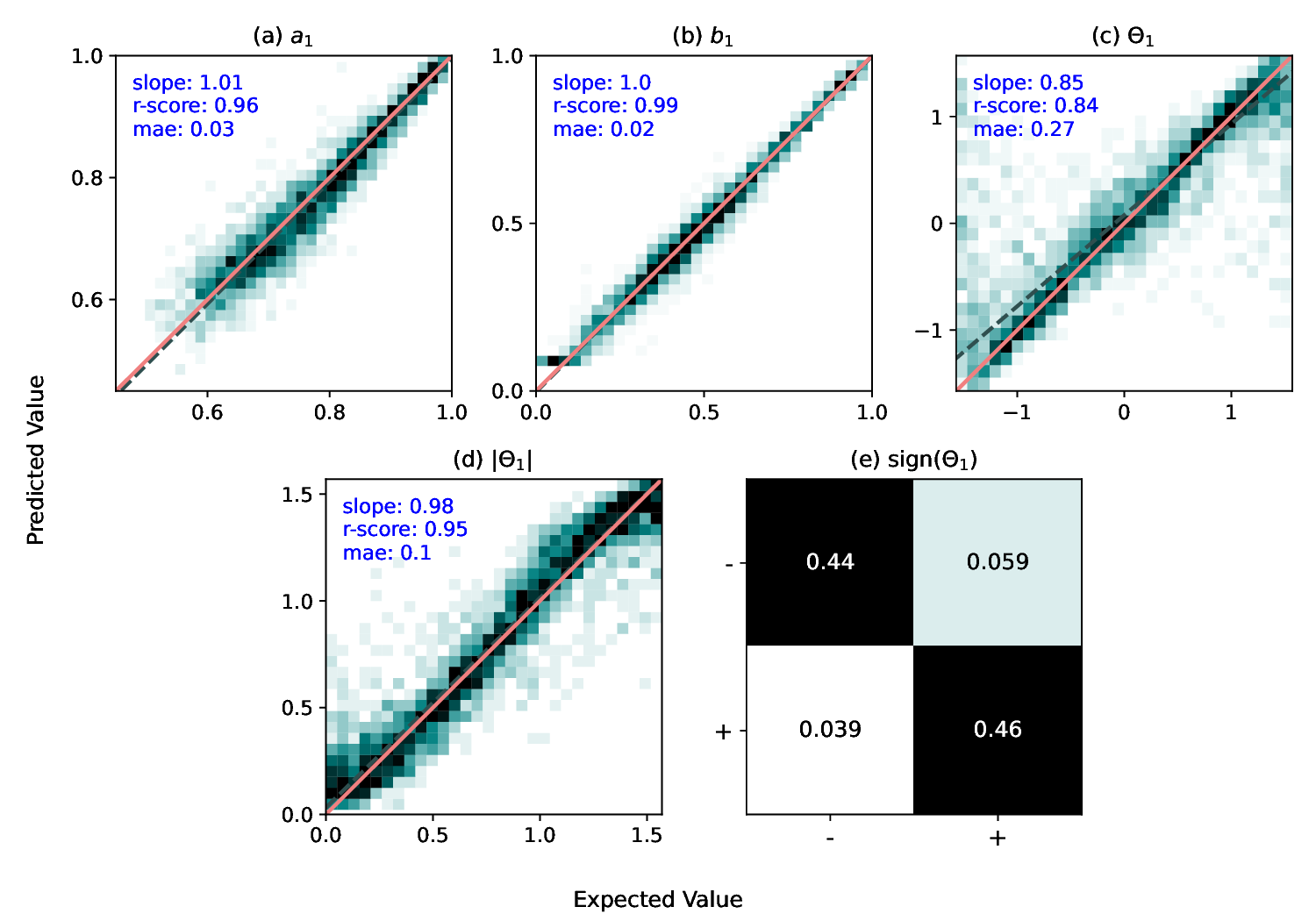}
    \caption{Expected vs. predicted values of the first order Fourier coefficients. The panels (a), (b), (c) show the expected vs. predicted value of $a_1$, $b_1$ and $\Theta_1$ respectively. The solid line shows the 1-1 relation. The dashed line represents the slope of the best fit lines between predicted and expected values. The r-score represents the Pearson correlation coefficient and mae represents the mean absolute error in prediction of test sample.  Panel (d) shows the training results for $|\Theta_1|$. Panel (e) shows the confusion matrix for $sign(\Theta_1)$.
    \label{fig:fr_1_train}}
\end{figure*}

\subsection{Complexity Training} \label{subsec:complexity}
  The complexity metric ($\mathcal{C}$) can itself be thought of as a one-dimensional vector encoding the shape parameter space. We train the neural network to take the light curve as input and predict the complexity value of the corresponding shape. The predicted and expected $\mathcal{C}$ for shapes in the test sample is shown in Fig. \ref{fig:complex_train}. We see that the predicted and expected values represent a moderately strong correlation, with an r-score of $0.73$, and a slope of $0.8$. This shows that the ML model is able to predict the complexity parameter decently. The lower $\mathcal{C}$ values are predicted better than higher $\mathcal{C}$ values. In the range $0.4-0.8$, we observe a tendency of the neural network to underpredict the complexity values, leading to outliers, and a lower slope estimate. This shows that certain features that lead to high complexity may not be embedded in the light curve. Nevertheless, this exercise demonstrates the ability of the NN to extract shape information from a light curve. In the next subsection, we repeat the same exercise for the first-order Fourier coefficients. 

\subsection{The Fourier First Order} \label{subsec:first_order}

  \begin{figure}
    \includegraphics[width=1\linewidth]{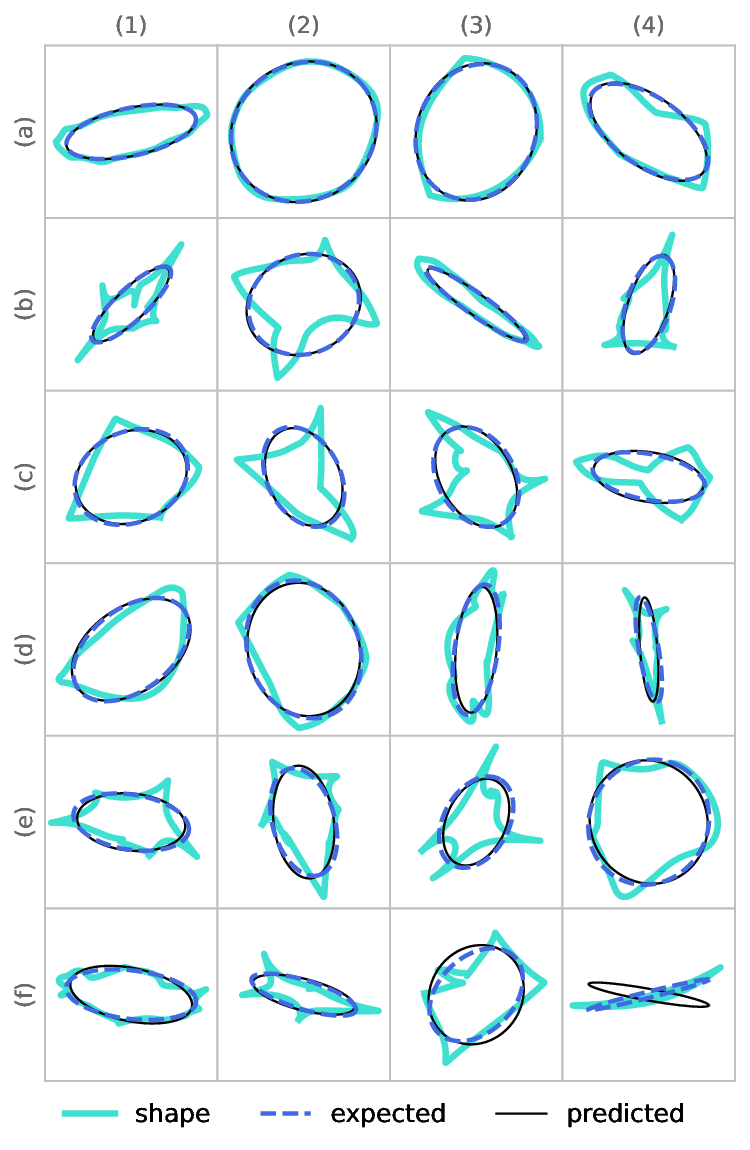}
    \caption{Examples of shapes from the test sample along with their expected first order Fourier ellipse (dashed lines) is shown. The Fourier ellipse drawn from the predicted values of the first order Fourier coefficients are shown as solid black line.
    \label{fig:fr_1_result}}
  \end{figure}

  The Fourier First order coefficients $(a_1, b_1, \Theta_1)$ form an ellipse, which gives us a measure of the magnitude and the overall orientation of the anisotropic features in a given shape. We train neural networks to take light curve input and predict the values of $a_1, b_1, \Theta_1$. We perform hyperparameter tuning to obtain the best fit model, which is given in Table \ref{tab:model-selection}. The results of the training are also shown in Fig. \ref{fig:fr_1_train}. We get $r=0.96, m=1.01$ for $a_1$ and $r=0.99, m=1$ for $b_1$ (panel (a) and (b)). This shows that the NN is able to accurately predict the value of $a_1$ and $b_1$ from the light curve input. Therefore, the eccentricity of a shape is embedded in the transit light curve, and a highly eccentric shape is distinguishable from a more circular shape. The mean absolute error (MAE) for prediction of $a_1$ is $0.03$, which corresponds to an error of $0.009 R_{st}$. The  MAE for $b_1$ is $0.02$, or $0.006 R_{st}$.

  Fig. \ref{fig:fr_1_train}(c) shows the results of training the orientation of the ellipse $\Theta_1$. Here, we see that a number of points lie on the outliers, where the expected value of $\Theta_1$ is $-\pi/2$ or $\pi/2$. In addition, we get $r=0.84, m=0.85$ and MAE$=0.027 rad$. Due to the impact of flip degeneracy, the outliers show predictions of correct magnitude, but incorrect sign. Also, at the boundary, $\pi/2$ and $-\pi/2$ correspond to the same orientation angle, but lead to large values of loss in during training. To improve this, we train $|\Theta_1|$ separately and design a classifier network to learn $sign(\Theta_1)$. We obtain a significantly better training result for $|\Theta_1|$ with $r=0.95, m=0.98$ and MAE=$=0.1$  as shown in Fig. \ref{fig:fr_1_train}(d and e). Separately trained, the sign of $\Theta_1$ also shows a classification accuracy of $0.90$. Combining the separately trained $|\Theta_1|$ and $sign(\Theta_1)$ leads to an overall improvement in the prediction of $\Theta_1$, with a MAE of $0.16\,rad$, when compared to the earlier case, with MAE$=0.27\, rad$. 

  A proper learning of the Fourier first order implies that under the given assumptions, it is possible to retrieve the magnitude, eccentricity and orientation of the overall anisotropy of an arbitrary geometry. We call this the `effective ellipse' of the shape. This is depicted in Fig. \ref{fig:fr_1_result}. Example shapes are shown along with the ellipses formed by the first order Fourier coefficients. The predicted ellipses are shown as solid lines. The shapes shown are sampled in increasing order of reconstruction error, between the predicted and expected ellipse. We see that the predicted ellipses match closely with the expected ellipses. The last shape: 4-f shows a significant deviation between the two ellipses, however, the network has likely predicted the ellipse flipped about the x-axis, which is a degenerate scenario. This exercise demonstrates the successful embedding of Fourier first order coefficients in the transit light curve.


\section{Predictions of Undulations over Effective Ellipse} \label{sec:higher_order}

Predictions of further higher order Fourier coefficients will lead to the predictions of distortions over the effective ellipse. So we repeat this exercise for the higher order Fourier coefficients.

\subsection{higher order a and b}

  The higher order $a_n$ govern the scale of the perturbations to the effective ellipse. Since all generated shapes show a logarithmic decline in $a$ for subsequently higher orders, we first take $log(a_n)$ and normalize the values between 0-1. For demonstration, we train the neural network: D:128-128-64-64-64-32 to learn the parameters $a_n \forall n \in [2, 20]$.

  \begin{figure}
    \includegraphics[width=1\linewidth]{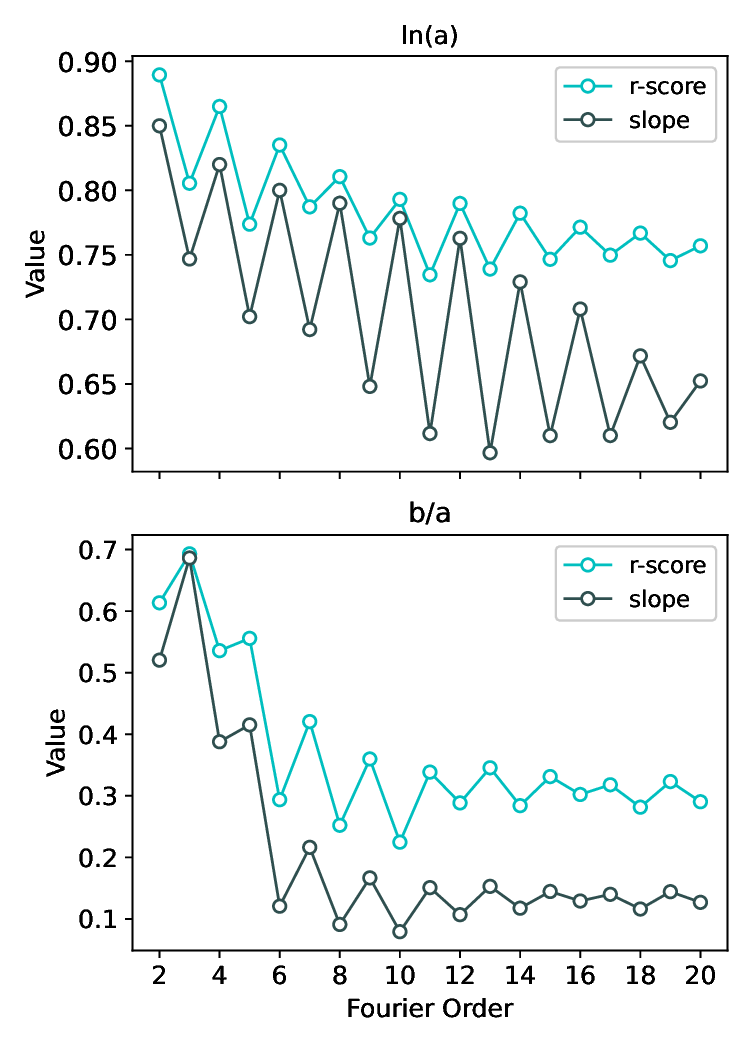}
    \caption{Training results of higher order $a, b$. The top panel shows the $r$ (Pearson correlation) and $m$ (slope of best fit line) between predicted and expected values of $ln(a_n)$ for higher order coefficients. The bottom panel shows the $r, m$ values for higher order $b_n/a_n$. 
    \label{fig:fr_a_b_train}}
  \end{figure}

  The results of the training is shown in the top panel of Fig. \ref{fig:fr_a_b_train}. The figure shows the r-score and slope between the predicted and expected parameters for each Fourier order. We see that the values of r-score and slope are lower for subsequently higher orders as compared to the first order. Both r-score and slope decline over higher orders and saturate at $r = 0.78, m = 0.65$. We also observe an alternating effect between odd and even orders. This is attributed to the variation in values seen in the training sample (see Fig. \ref{fig:par_dist}). The orders for which there is a greater variation in values show a higher score in training. Even then, the r-score value remains high for very high order Fourier coefficients, showing that the NN is able to predict this parameter up to very high orders. 

  For higher orders, we train for the parameter $b/a$. We train the NN D:128-128-64-64-64-32 to learn higher order $b/a$. The results of the training is shown in the bottom panel of Fig. \ref{fig:fr_a_b_train}. The figure shows the r-score and slope between the predicted and expected parameters for each Fourier order. We see a moderately strong r-score for Fourier orders up to 5, after which the r-score drops sharply. This shows that the NN is unable to predict eccentricity for very high Fourier orders. Therefore, we perform hyperparameter tuning for $b/a$ parameter only up to Fourier order 5. The best fit model and the resulting r-score and slope are mentioned in Table \ref{tab:model-selection}.

  \begin{figure}
    \centering
    \includegraphics[width=1\linewidth]{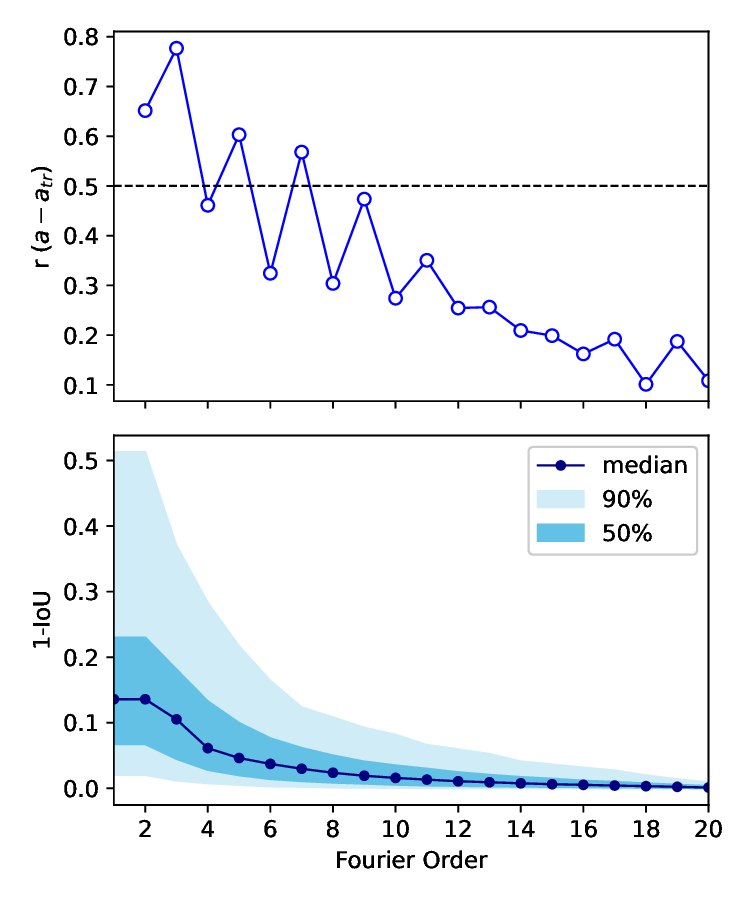}
    \caption{Pearson correlation of $a_n - a_{trend}$ between expected and predicted values of $a_n$ is shown in the top panel. The horizontal line marks $r=0.5$. The bottom panel shows the statistics of reconstruction error, if values for a Fourier order higher than a given order are replaced by $a_{trend}$ for all shapes in the training sample. The markers represent the median, and the solid contours enclose $50\%, 90\%$ of the data for a given order.}
    \label{fig:fr_jitter}
  \end{figure}  

  The logarithmic trend observed in $a$ in the training samples and the saturation of the r-score for very high orders in $a$ suggest that the NN may be predicting values from the trend of $a$ rather than successfully learning individual values. To explore this possibility, we fit equation \ref{eq:trend} to the $a$ values for each shape. The $a$ value for different orders can be calculated using two parameters $\Lambda, \gamma$; we call these values $a_{trend}$. For each shape, we subtract $a_{trend}$ from both the expected as well as predicted values of $a_n$. We calculate the r-score for expected and predicted values of $a_n - a_{trend}$. This is shown in Fig. \ref{fig:fr_jitter} (top panel). We find a moderately strong correlation for orders 2 - 5, which drops to very low values for higher orders.  This lack of correlation shows that for higher order Fourier components, the NN is unable to learn the individual parameter values with any better accuracy than defined by the logarithmic trend. Therefore, it is unnecessary to train the NN for individual components for orders higher than 5, and instead it can be trained to learn the trend of $a$, given by parameters $\gamma_a, \Lambda_a$.

  In order to understand the effect of substitution of $a_n$ with $a_{trend}$, we perform the following exercise. For a given Fourier order $n$, we preserve original values of $a$ up to order $n-1$, and for orders greater than that, we replace with values obtained from $a_{trend}$. Then we compare the reconstruction error between the original shape, and the shape generated by this substitution. We do this for different values of $n$ for all shapes in the training sample. The results of this exercise are shown in the bottom panel of Fig. \ref{fig:fr_jitter}. The median reconstruction error and the 50, 90 quantile of the distribution for each order is shown. For order 1, 2 all the values except $a_1$ have been replaced by $a_{trend}$. So the reconstruction error is high. As we go to higher orders, the reconstruction error diminishes. We observe from this figure that beyond Fourier orders 5, such a substitution will lead to low reconstruction error, which means a minimal deviation from the original shape. If we replace $a$ with $a_{trend}$, starting from Fourier order 5, $75\%$ shapes will have $IoU>0.9$, $90\%$ of shapes will have $IoU>0.8$.

\begin{figure}
    \centering
    \includegraphics[width=1\linewidth]{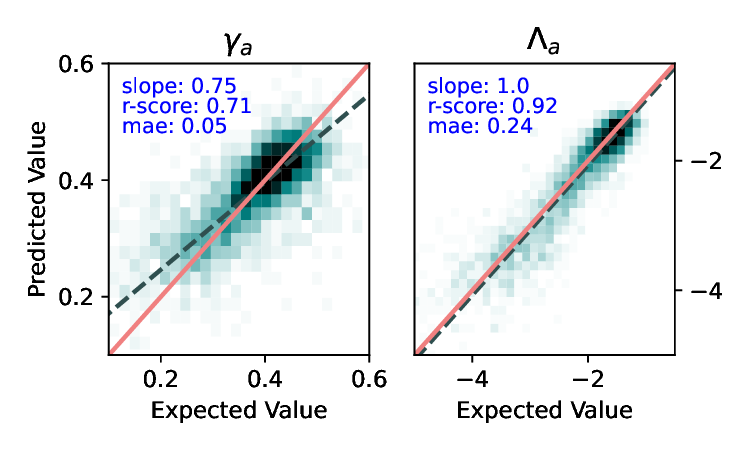}
    \caption{Expected vs Predicted values of logarithmic trend parameters in Fourier $a$. The left and right panels show the training results for $\gamma_a$ and $\Lambda_a$ respectively. The  slope and r-score values represent the slope of the best fit line (shown as a dashed line) and the Pearson correlation coefficient for each case. The solid line represents a 1-1 relation.}
    \label{fig:fr_trend_train}
  \end{figure}

  We perform hyperparameter tuning to obtain the best performing network for parameters $a_2 - a_5$, as mentioned in Table \ref{tab:model-selection}. For further higher orders, we train neural networks to learn the parameters $\Lambda_a, \gamma_a$ for $a$. The best performing networks are mentioned in Table \ref{tab:model-selection}. The results of the training are shown in Fig. \ref{fig:fr_trend_train}. We see that the NN learns $\Lambda_a$ with high accuracy, as seen by $r=0.95$ and $m=1.0$. The $\gamma_a$ parameter is comparatively more challenging to learn. However, it still shows good correlation, with $r=0.81$ and $m=0.75$. 

\subsection{higher order $\Theta$ and $\phi$}

  In this subsection we repeat the same exercise as above for parameters $\Theta_n$ and $\phi_n$. As observed in the case of $\Theta_1$, due to flip degeneracy, NN model may confuse $\Theta$ and $-\Theta$ for a given order. Also, an incorrect prediction of sign would be heavily penalized by the network, instead, the learning of the magnitude is likely to lead to a greater reconstruction loss. Therefore we train the magnitude and sign of $\Theta_n$ and $\phi_n$ using separate networks. 

  \begin{figure}
    \includegraphics[width=1\linewidth]{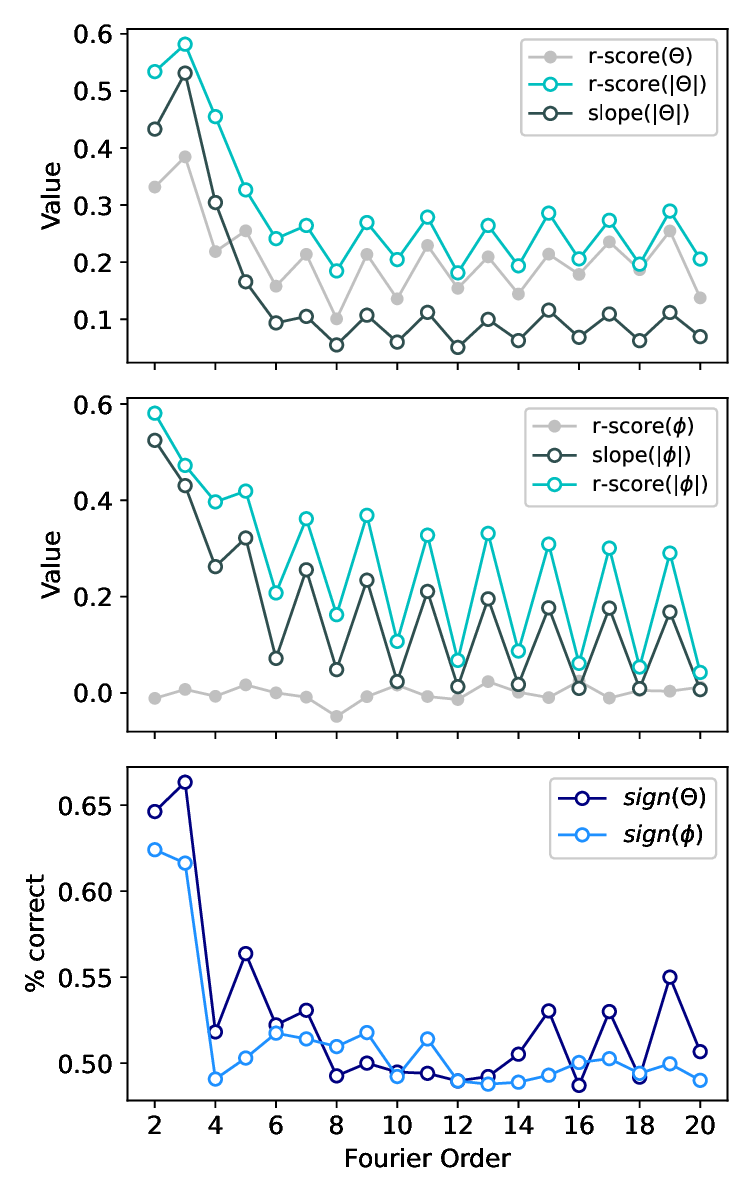}
    \caption{Training results of higher order $\Theta, \phi$. The top panel shows the r-score and slope for higher order $|\Theta|$ The r-score for training of $\Theta$ is also shown in gray. The middle panel shows the r-score and slope for $|\phi|$ along with r-score for $\phi$ The bottom panel shows the accuracy of binary classification on the sign of $\Theta$ and $\phi$.
    \label{fig:fr_r_th_ho}}
  \end{figure}

  We train for parameter $|\Theta_n|$ and $|\phi_n|$ using model D:128-128-64-64-64-32. The results of the training are shown in Fig. \ref{fig:fr_r_th_ho}. The top panel shows the r-score and slope between predicted and expected values of $|\Theta|$ in the test sample. We observe a moderate correlation for orders 2-4 after which the r-score drops to $0.25$. For demonstration, the r-score obtained by training $\Theta$, without the sign separation is also shown. The r-score for this case is significantly lower, especially in orders 2-4. In addition, the separation of magnitude and sign improves the overall MAE of the $\Theta$ parameter. We obtain a MAE of $0.54, 0.48$ in orders 2,3 for the sign separated network as compared to an MAE of $0.66, 0.63$ for the combined network. The MAE of orders higher than 4 are comparable. This along with the low r-score for higher orders suggest that the information embedded in the light curves is limited for these orders.

  The middle panel shows the r-score and slope for $|\phi|$. Here also we see a moderate r-score $(\sim0.6)$ for Fourier orders 2-3, after which the r-score decreases. We also see strong variations in r-score between odd and even orders. This is attributed to the variance in the training set that is observed in parameter $\phi$ (see Fig. \ref{fig:par_dist} last panel). The r-score obtained by training $\phi$, without the sign separation is shown. We see that that without the sign separation the NN completely fails to learn higher order parameters in $\phi$. In addition, we obtain an MAE $1.15, 0.89$ for orders 2 and 3 using the sign separated network, as compared to MAE $1.57, 0.96$ using combined network.

  \begin{figure*}
    \includegraphics[width=1\linewidth]{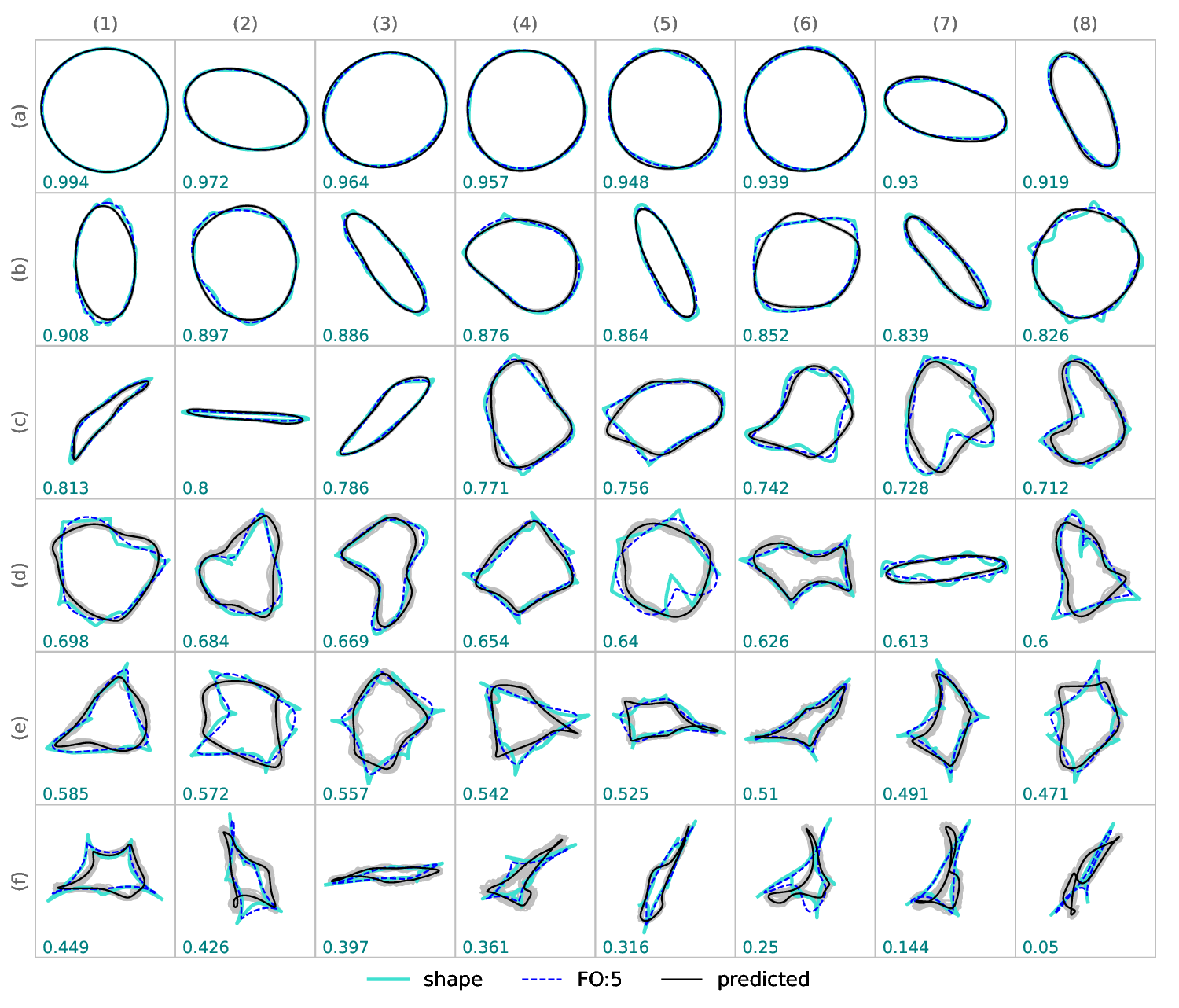}
    \caption{Original vs Reconstructed shape based on the elliptical Fourier components predicted by the neural network. The blue lines represent the original shape, whereas the dashed lines represent the expected shape formed by $5$ Fourier order (labeled FO:5) components of the original shape. The gray curves represent a set of $100$ randomly sampled possible shapes based on the NN predictions. The solid black line represents the mean of these predicted shapes. The values next to each shape show the IoU between the original and the mean of the predicted shapes.
    \label{fig:combine_rec}}
  \end{figure*}

  The parameters $\Theta, \phi$ are the most challenging to learn, as we observe from the low r-scores in higher order training. Besides, the relevance of $\Theta_n, \phi_n$ for a Fourier order is influenced by the scale of $a$ and eccentricity (proxy by $b/a$) of that given order. If the $a$ parameter has a lower value, then the effect of $\Theta, \phi$ on the shape would be lower as well. From the results, we consider Fourier order up to 4 for hyperparameter tuning. The best model and its performance is listed in Table \ref{tab:model-selection}. 

  The last panel of Fig. \ref{fig:fr_r_th_ho} shows the training results of a binary classifier trained to learn the sign of $\Theta, \phi$. At best, for Fourier orders 2-3, the classifier is able to give validation accuracy of $60-65\%$. For subsequent orders, the accuracy goes to $50\%$, which shows that the NN is unable to extract the sign information from the light curve for these higher orders. We perform hyperparameter tuning for orders up to 5. In this case, hyperparameter tuning leads to significant improvements in the model performance. The best performing models and their accuracy are given in Table \ref{tab:model-selection}. 
  
\subsection{combined reconstruction}
  From the parameters predicted by the neural network, we can create a reconstruction of the shape. Each parameter up to Fourier order 5 is trained individually, and the best performing model is selected (from Table \ref{tab:model-selection}). For further higher orders, the $a$ values are taken from the trend of $a$, and the rest of the parameters are randomly sampled to create a set of possible shapes from the given predictions. Fig. \ref{fig:combine_rec} shows a sample of shapes and their reconstructions. The solid blue lines show the original shape, the dashed line shows the shape that is obtained if only 5 Fourier orders are considered. We can see visually that shapes can be approximated to a good extent with 5 Fourier orders (except for some spikes in the more complex shapes). The gray lines represent the set of shapes obtained by randomly sampling the parameters that the NN is unable to predict. The solid black line represents the mean of this sample.

  The shapes have been arranged in a decreasing order of IoU between the mean reconstructed shape and the original shape. The top rows show the shapes that have been reconstructed well, as represented by the high values of IoU. For the shapes that have been reconstructed well, the width of the gray region is also smaller, i.e the deviations due to random samplings in the higher order parameters do not impact the shape to a great degree. For these shapes, most of the shape features are embedded in the lowest Fourier orders. 

  As we go to the bottom rows, The IoU decreases. Many of the shapes in the bottom row are complex shapes with small area. In this case, the transit depth is lower, which may contribute to the poorer reconstruction. Correspondingly, the width of the gray region increases. The deviations due to random samplings in the higher order create more significant impact in the shape reconstruction. For shapes f6 - f8 the IoU values are quite small. Despite that, we can see that the reconstructed shape captures some aspects of the anisotropy. The median IoU of reconstruction obtained over the test sample is $0.67$, and $76\%$ of the test sample is reconstructed with an $IoU>0.5$. For most of the shapes, we see that the NN is able to capture the general anisotropy of the original shape. This gives us visual cues towards the extent of reconstruction that is possible from the transit light curves.


\section{Discussion} 
\label{sec:discuss}

We combine all the NN predictions of parameters to generate reconstructed shapes for the training and test sample. Then we calculate the reconstruction error between the original shape and the reconstructed shape. 

A demonstration of the influence of individual parameters on the shape reconstruction is the correlation between error in parameter prediction, and the overall reconstruction of the shape. We calculate the r-score between the MAE of the predicted parameters and the net reconstruction error of the shape. The mean r-score for $a$ is $0.38$ for the first 4 Fourier orders, and $0.36$ for $b$, which implies a weak correlation between the error in predicted parameter and the overall reconstruction error. On the other hand, performing a similar calculation for $\Theta$ gives us an r-score $-0.04$, and $-0.02$ for $\phi$. This highlights the challenges in learning of the $\Theta, \phi$ parameters, and their complex influence in the reconstruction of the shape. In the following subsections, we examine such statistical relations of the net reconstruction error to various parameters, such as the complexity and convexity of individual shapes.

\subsection{Relations between Predicted Complexity and Shape}

  \begin{figure}
    \includegraphics[width=1\linewidth]{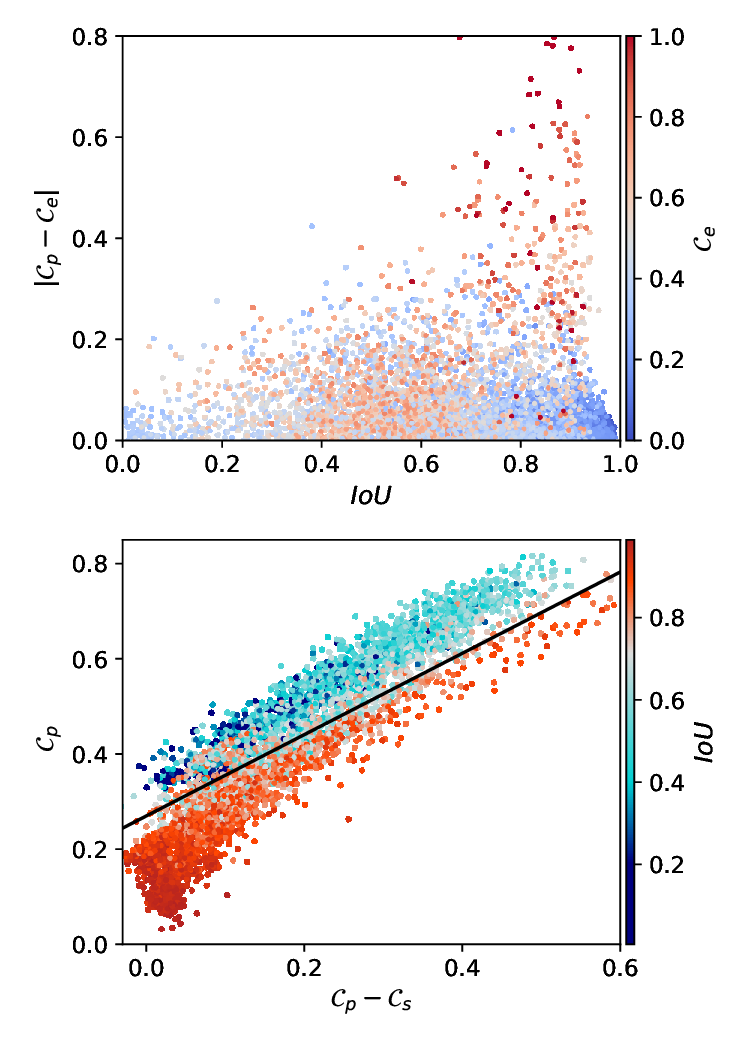}
    \caption{The top panel shows the $IoU$ between original and reconstructed shape in x-axis, and the error between the predicted and expected complexity ($\mathcal{C}_p-\mathcal{C}_e$) on the y-axis. The color bar shows the expected complexity ($\mathcal{C}_e$). The bottom panel shows the error between the predicted complexity and the complexity of reconstructed shape on the x-axis ($\mathcal{C}_p-\mathcal{C}_s$), and the predicted complexity($\mathcal{C}_p$) on the y-axis. The color bar shows the $IoU$ between the shape and its reconstruction.
    \label{fig:comp_rec}}
  \end{figure}

  In section \ref{subsec:complexity}, we trained a NN to learn the complexity parameter ($\mathcal{C}_e$) of the shape from light curve input. The complexity predicted from this network is $\mathcal{C}_p$. The same metric can be applied to the reconstructed shapes to obtain a complexity value for the reconstructed shape, denoted as $\mathcal{C}_s$. Ideally, for a shape that has been reconstructed accurately, the reconstructed shape complexity ($\mathcal{C}_s$) should match the predicted complexity ($\mathcal{C}_p$) and expected complexity ($\mathcal{C}_e$).

  Fig. \ref{fig:comp_rec} compares the performance of the complexity prediction and shape reconstruction. The top panel shows the $IoU$ between the original and reconstructed shape vs. the difference in expected and predicted complexity ($|\mathcal{C}_p-\mathcal{C}_e|$). We see that for a majority of shapes the complexity prediction is fairly accurate ($90\%$ samples with error$<0.2$), even when the $IoU$ has low values (poor reconstruction). The scatter-plot is colored by the expected complexity of the original shape ($\mathcal{C}_e$). For the lowest complexity values, both $|\mathcal{C}_p-\mathcal{C}_e|$ is close to $0$ and the $IoU$ is close to $1$. This shows that the reconstruction of the shape as well as the complexity prediction is accurate for these shapes. For complexities in the range $0.3-0.7$, we see the maximum variation in $IoU$. However, the complexity prediction error is low. 

  For a small fraction of shapes we see that $|\mathcal{C}_p-\mathcal{C}_e|$ is high. However, for these shapes the $IoU$ is high. This means that the predicted shape, even with much lower complexity than expected, is able to represent the shape with good accuracy. This is indicative of shapes where the high value of complexity arises due to localized distortions to an otherwise smooth shape, and not a globalized anisotropy. For example, in Fig. \ref{fig:sh_c_eg}, shapes c-5 and d-7 have very different complexity values, but both shapes have very small distortions over an otherwise elliptical shape. Therefore, prediction of an ellipse would give rise to very low reconstruction loss, despite high values of complexity.

  The bottom panel of Fig. \ref{fig:comp_rec} shows the difference between the complexity predicted by the NN and the complexity of the reconstructed shape ($\mathcal{C}_p-\mathcal{C}_s$) on the x-axis, and the predicted complexity ($\mathcal{C}_p$) on the y-axis. The scatter-plot is colored by the $IoU$ between the original and reconstructed shape. With increase in $\mathcal{C}_p$, $\mathcal{C}_p-\mathcal{C}_s$ increases. This is expected, since for high complexity shapes, the reconstructed shape may not match the original complexity. However, we see clear regions where reconstruction loss is low ($IoU$ is high), and regions where $IoU$ is low. Now for an unknown light curve, both $\mathcal{C}_p-\mathcal{C}_s$ and $\mathcal{C}_p$ can be obtained from the NN output. Based on the location of the sample on this plane, we can infer the extent of reconstruction loss of shape corresponding to a new light curve. 

  On the $\mathcal{C}_p-\mathcal{C}_s$,$\mathcal{C}_p$ plane we see clustering based on the $IoU$ values in the scatter-plot. An empirical linear decision boundary can be obtained based on the $IoU$ value. To identify the empirical decision boundary, we first select a threshold value for $IoU$. Based on this threshold, if $IoU$ is higher or lower than this threshold, we label the scatter-point as 1 or 0. Then we fit a support vector classifier \citep{SVC}\footnote{\textsc{Scikit-learn} implementation of SVC is used}. The fitted classifier, divides the $\mathcal{C}_p-\mathcal{C}_s$, $\mathcal{C}_p$ plane into two regions separated by linear boundary. For the classifier, the input features are $\mathcal{C}_p-\mathcal{C}_s$ and $\mathcal{C}_p$ and the target label is 0 or 1 based on condition that $IoU>$ a selected threshold. The classifier is trained and validated on the training-dataset (24000 samples) and accuracy is computed on the held-out test set (6000 samples). We achieve the maximum classification accuracy of $86.7\pm0.3$\% for $IoU$ threshold at 0.72. (cross validation and threshold calculation is discussed in Appendix \ref{apdx:iou}). The solid line in Fig. \ref{fig:comp_rec} shows the linear decision boundary, corresponding to $IoU=0.72$. For a new light curve, the two set of complexities ($\mathcal{C}_p, \mathcal{C}_s$) can be independently computed. This classifier then allows us to assert the quality of reconstruction by estimating its IoU above or below the threshold 0.72. If the point lies above the decision line on the $\mathcal{C}_p-\mathcal{C}_s$,$\mathcal{C}_p$ plane, then with $\sim 87\%$ accuracy we can say that $IoU$ of the shape is above $0.72$.

\subsection{Implications of Shape Convexity}

  \begin{figure}
    \includegraphics[width=1\linewidth]{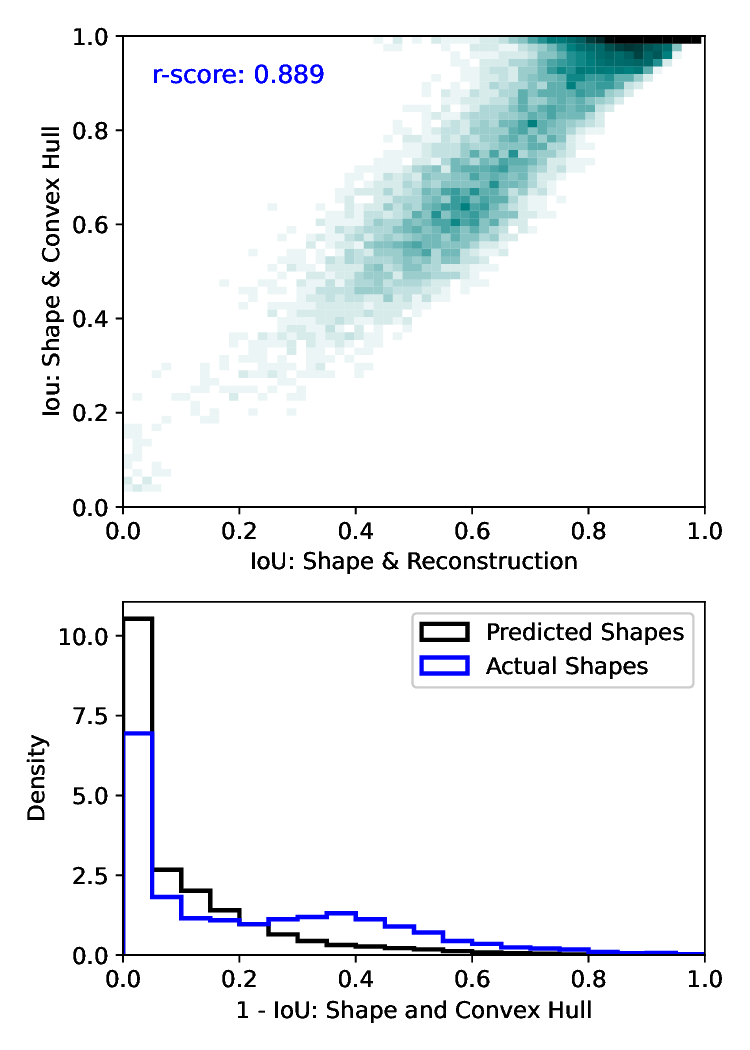}
    \caption{Relations between the non-convex nature of a shape and its reconstruction. The top panel shows the $IoU$ between the original and reconstructed shape, and the $IoU$ between the shape and its convex hull on the y axis. The r-score of the data is given. The bottom panel shows the histogram of $1-IoU$ between a shape and its convex hull. This is shown for both the original as well as the reconstructed shape.
    \label{fig:convex_hull}}
  \end{figure}

  From reconstructed shapes shown in Fig. \ref{fig:combine_rec}, we see some large deviations in shapes that have large concave regions (panels e5, c7, d2). To investigate this further, we generate a convex hull of the expected shape and the predicted mean shape. Then we calculate the $IoU$ between the shapes and their convex hull. The smaller the $IoU$, the more non-convex the shape. 

  The top panel in Fig. \ref{fig:convex_hull}, shows the $IoU$ between the original and reconstructed shape, vs. the $IoU$ between the shape and its convex hull. We see a significant correlation, with an r-score: $0.89$. This correlation indicates that the more non-convex the shape, the greater is the reconstruction error. This may indicate that NN is unable to learn non-convex features from the light curve input. The bottom panel shows the histogram of the difference between the shape and its convex hull ($1-IoU$) for the original, as well as the predicted shape. The histogram for the actual shapes show significant number of non-convex shapes in our training sample ($1-IoU:0.2-0.6$). However the histogram for the predicted shapes is skewed towards $0$. This analysis indicates that the embedding of non-convexity in the transit light curve is potentially limited. \cite{kaasalainenI_2001, kaasalainenII_2001} arrives at a similar conclusion on asteroid shape inversion using occultation light curves. Concave shapes are likely to be degenerate with equivalent convex shapes. 

\subsection{Orientation Dependence of Concave Shapes}

  \begin{figure*}
    \includegraphics[width=1\linewidth]{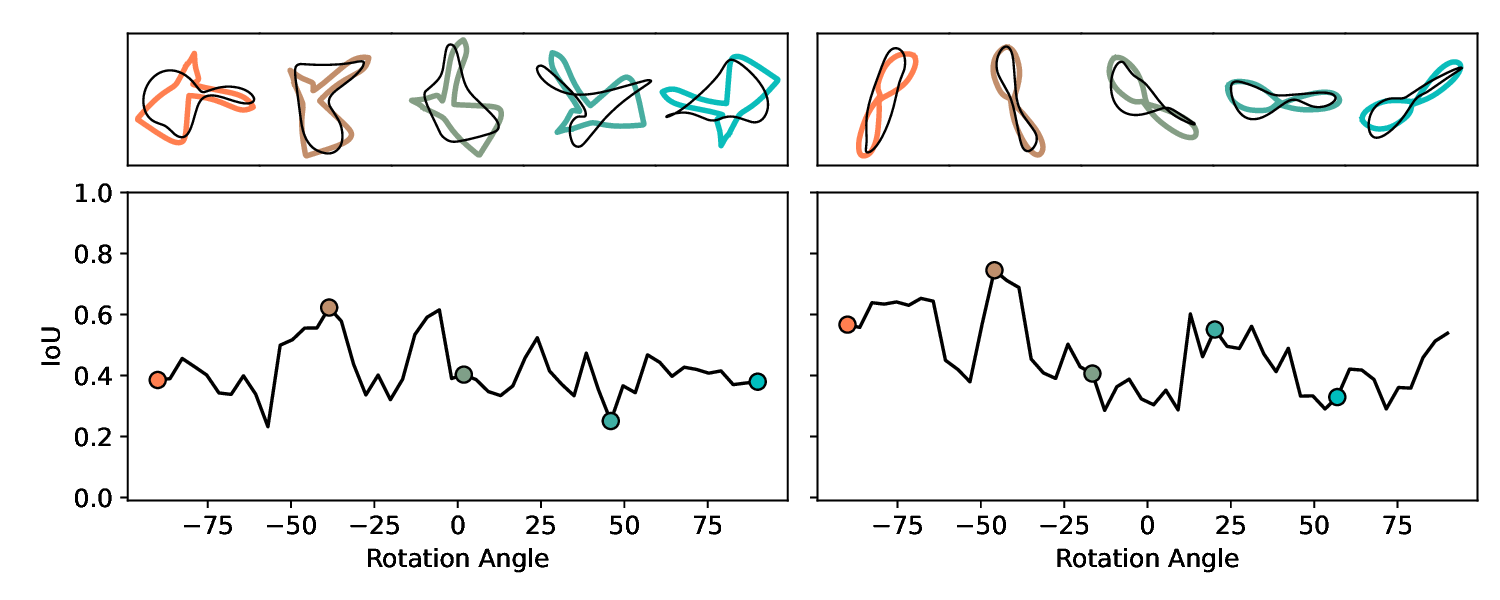}
    \caption{Demonstration of the effect of shape orientation in reconstruction for a non-convex shape. The top panels show a shape in 5 different orientations, and the reconstructed shape as a solid black line. The bottom panels shows the $IoU$ between original and reconstructed shape for different rotation angles. The rotation angles of the shapes in the top panel are marked.
    \label{fig:oriented_concave}}
  \end{figure*}

  The analysis of the previous subsection indicates that the NN is unable to learn non-convex features from the light curve. In this subsection, we investigate the effect of the orientation of the non-convex shape in its reconstruction. We select two examples of non-convex shapes from our training sample. We rotate these shapes from $-90^{\circ}\rightarrow90^{\circ}$ angle. We generate $50$ rotated shapes, for which we simulate the transit light curve using \yuti. Then we estimate the Fourier parameters using the neural networks, and use them to reconstruct the shape. 

  Fig. \ref{fig:oriented_concave} shows the result of this exercise for the two selected shapes. The top panels show $5$ different orientations of the same shape, and the reconstructed shape in black. Both the chosen shapes have non-convex features, but the reconstructed shape varies for different orientations. In fact, in some orientations, the NN is able to capture some aspects of the concave features. The bottom panel shows the variation of $IoU$ between the original and the reconstructed shape for different orientation angles. The $IoU$ shows significant variations across different angles. The angles having the maximum $IoU$ visually show a better reconstruction. This exercise shows that while non-convex features may be poorly embedded in the light curve, but the orientation of the non-convex shape can bring about significant differences in reconstruction. 

\subsection{Caveats}
  The analysis done in this work shows the extent of shape features embedded in a transit light curve. A number of simplifying assumptions have been made for this analysis. The transit parameters are fixed, and no limb darkening of the star has been considered. For instance, \cite{oblatedeg_barnes_2003}, explores the case of distinguishing between an oblate and spherical planet and shows that variation in size and impact parameter during fitting can diminish the effect of distortion due to the oblateness. On the other hand, at high impact parameters, the difference between the oblate light curve and its best-fit spherical light curve is maximized. Similarly \cite{yuti_bhowmick_2024} compares the light curve of a transiting planet with that of a hypothetical `dyson disk'. It is shown that the deviations in the transit light curve are diminished during fitting by assuming higher limb darkening coefficients. \cite{exoringInv_arkhypov_2021} demonstrates the merits of independent measurements of orbital distance and limb darkening to better constrain non-spherical shapes. Therefore, understanding the impact of transit parameters and limb darkening is crucial for understanding light curves for realistic transiting systems.

  Observations of transiting systems will also contain instrumental and random noise. In order to apply this method to real transit observations, the NN must be trained on noise-added light curves. The effect of noise on the performance of the NN is expected to be complex. It depends on the transit parameters, such as size, orbital distance and impact parameter. It also depends on the impact of a given Fourier coefficient on the transit light curve of a particular shape. A preliminary analysis is given in Appendix. \ref{apdx:lc-noise}. The performance values shown in this current work represent an upper limit on light curve inversion performance.
  
  The current method of 2D shape generation and analysis with \bez curves and elliptical Fourier descriptors does not consider shapes with holes, such as silhouettes of rings. Also, certain realistic systems such as comets with evaporating tails are expected to show silhouettes which change during the transit. A more general analysis will also take into consideration full 3D arbitrary objects and the effects of rotation. Currently, this remains as a limitation of this analysis. We will explore these ideas in our upcoming works.

  Nevertheless, this work demonstrates a method to extract shape information from light curves. The simplified 2D-shape transit has potential applications for certain systems such as flybys of objects around stars, or transits of objects with distant orbits.

\section{Conclusion} \label{sec:conclusion}
 
  We investigate the extent of geometric information that can be extracted from a photometric transit light curve. This problem is ill-posed as many different shapes can produce indistinguishable light curves. To address this, we generate a library of arbitrary 2D shapes and their corresponding simulated transit light curves using \yuti. We train a neural network to predict the complexity of the 2D shape directly from the light curve. With this parameter, we obtain an immediate quantitative flag for departures from simple shapes. For reconstruction of complex shapes, we represent each shape as an elliptical Fourier decomposition. We use neural networks to predict different order Fourier coefficients of the 2D shapes with light curve given as the input, thereby training an inverse model that predicts the underlying shape from the light curve.

  The neural networks predict the first-order Fourier coefficients with high accuracy. The effective ellipse formed by the first order gives us an estimate of the scale, eccentricity and orientation of the overall anisotropy in the shape, which is well embedded in the transit light curve. Accurate predictions of the effective ellipse from observed light curves can help distinguish complex natural phenomena, such as tidally or rotationally distorted planets \citep{tide_saxena_2015, oblate_berardo_2022}, which have elliptical projections, or extended atmospheres of close-in planets \citep{hj_zhilkin_2020, radiotr_selhorst_2020}. A highly eccentric ellipse with an unusual orientation can be indicative of a more complex or exotic phenomena.

  Higher-order Fourier terms are predicted with progressively lower fidelity. Beyond Fourier order 5, the parameters $b$ (semi-minor axis), $\Theta$ (orientation), $\phi$ (starting location) of the component ellipse cannot be retrieved by the neural network, and the correlation coefficients fall substantially. The systematic trend in $a_n$(semi-major axis) for higher-orders is learnt by the NN, which can help predict $a_n$ for very high Fourier orders. Predictions up to Fourier order 5 give us a good reconstruction of the shape for most of the shapes. This shows that some very asymmetric and complex features can be predicted from the transit light curve. The shapes predicted by the neural network are found to be more convex, highlighting the degeneracies associated with prediction of concave features from transit light curves. Reconstruction of concave features also depends on the orientation of the shape.

  Predictions of complexity parameter when combined with shape reconstruction provide clues to the extent of reconstruction of a shape. The comparison of predicted complexity and complexity of reconstructed shape can indicate the extent of reconstruction. The shape reconstruction analysis presented in this work has the potential to unravel features of anomalous events such as tidally‑distorted hot Jupiters, disintegrating or evaporating planets, ringed worlds \citep{tide_saxena_2015, exocomet_etang_2022, wasp103b_barros_2022, evpl_curry_2024}, or megastructure candidates. Transit light curve inversion can unravel a host of geometric information from transit survey missions such as Kepler and TESS, as well as from future transit observations. The deep-learning inversion analysis presented in this work offers a powerful method for identifying and characterizing a broad spectrum of astrophysical objects that produce anomalous transit signatures, extending the diagnostic reach of photometric surveys beyond conventional exoplanet detection.

  \begin{acknowledgments}
     Acknowledgements : \textit{We thank Asif M Mandayapuram, Vibhuti Bhushan Jha and Dr. Munn Vinayak Shukla (SSD, SAC) for their discussions and comments on this work. We thank Dr. Mehul R Pandya (SESG, EPSA, SAC), Dr. Rashmi Sharma (EPSA, SAC) and Nilesh Desai, Director SAC for their support in this activity. Finally, we thank the reviewers for their valuable comments, which have greatly improved the quality of this manuscript.}

\textit{Softwares}: tensorflow \citep{tf_abadi_2015}, Scikit-learn \citep{sklearn_pedregosa_2011}
\end{acknowledgments}


\begin{appendix}
  \section{Conversion between Elliptical Fourier Descriptions}
  \label{apdx:fourier-conversion}

    Using elliptical Fourier descriptors (EFD) up to $N^{th}$ order, a 2D closed curve $(x, y)$ of $M$ boundary points \(\{(x_i,y_i)\}_{i=0}^{M-1}\) is parameterized as follows using $t$ as sampling parameter:

  \begin{equation}
      x(t) = \sum_{n=1}^N c^{n}_1 cos( 2n\pi t ) + c^{n}_2 sin(2n\pi t)
      \label{eq:fr_xa}
  \end{equation}

  \begin{equation}
      y(t) = \sum_{n=1}^N c^{n}_3 cos(2n\pi t) + c^{n}_4 sin(2n\pi t)
      \label{eq:fr_ya}
  \end{equation}

  where $c^{n}_1, c^{n}_2, c^{n}_3, c^{n}_4$ are Fourier coefficients for the $n^{th}$ order. The equations of $x(t)$ and $y(t)$ does not contain constant terms as all the shapes are shifted to origin such that centroid of the shape is at the origin. The coefficients are computed as:

  \begin{align}
    c_1^n &= \frac{1}{2\pi^2 n^2}\sum_{i=0}^{M-1}
        \frac{\Delta x_i}{\Delta s_i}
        \bigl[ \sin(2\pi n t_{i+1}) - \sin(2\pi n t_i) \bigr],\\[4pt]
    c_2^n &= \frac{1}{2\pi^2 n^2}\sum_{i=0}^{M-1}
        \frac{\Delta x_i}{\Delta s_i}
        \bigl[ \cos(2\pi n t_i) - \cos(2\pi n t_{i+1}) \bigr]
  \end{align}
    
  where, $\Delta s_i$ is the normalized cumulative sampling distance and given as:
  
  \begin{align}
    \Delta s_i &= \sqrt{(x_{i+1}-x_i)^2 + (y_{i+1}-y_i)^2},
  \end{align}
  
  \begin{align}
    s_i       &= \sum_{k=0}^{i-1}\Delta s_k,\quad\quad
                  s_0=0,\quad\quad
                 T = \sum_{k=0}^{M-1}\Delta s_k 
  \end{align}
  \begin{align}
    t_i       &= \frac{s_i}{\sum_{k=0}^{M-1}\Delta s_k},\quad\quad
                  0\le t_i < 1
  \end{align}

  Similar equations follow for $c_3^n$ and $c_4^n$. Fourier coefficients for each order can be converted to equivalent parameters of a rotated ellipse to obtain the following description: 

  \begin{align}
        \begin{bmatrix}
            x \\
            y
        \end{bmatrix} = \sum_{n=1}^N
        \begin{bmatrix}
            -sin\Theta_n & cos\Theta_n \\
            cos\Theta_n & sin\Theta_n \\
        \end{bmatrix} 
        \begin{bmatrix}
            a_n cos(2n\pi t + \phi_n) \\
            b_n sin(2n\pi t + \phi_n)
        \end{bmatrix}
  \end{align}

  Where $a_n$ is the semi-major axis, $b_n$ is the semi-minor axis, $\Theta_n$ is the angle of orientation of semi-major axis, and $\phi_n$ defines the starting point on the ellipse for a given Fourier order $n$. The Fourier coefficients between these two descriptions can be converted as given below:

  \begin{equation}
    \begin{split}
        a = \frac{1}{2} (\sqrt{(c_1+c_4)^2 + (c_3-c_2)^2} + \\ \sqrt{(c_2+c_3)^2 + (c_1-c_4)^2})
    \end{split}\label{eq:for_a}
  \end{equation}

  \begin{equation}
    \begin{split}
        b = \frac{1}{2}(\sqrt{(c_1+c_4)^2 + (c_3-c_2)^2} - \\ \sqrt{(c_2+c_3)^2 + (c_1-c_4)^2})
    \end{split}\label{eq:for_b}
  \end{equation}

  \begin{equation}
    tan\Theta = \frac{ac_4 - bc_1}{bc_3 + ac_2}
  \end{equation}

  \begin{equation}
    tan\phi = \frac{bc_1 - ac_4}{ac_3 + bc_2}\label{eq:for_phi}
  \end{equation}

  For a given order, the ellipse can be traversed either clockwise or anticlockwise, which is determined by the sign of $b$. From equation \ref{eq:for_b}, we see that the sign can be positive or negative depending on the first and second terms. $a$ varies between $(0,1)$, $b$ varies between $(-1, 1)$, $\Theta$ varies between $(-\pi/2, \pi/2)$ and $\phi$ can vary between $(-\pi, \pi)$. Equation \ref{eq:for_phi} has two solutions in the given range. To choose between these solutions we use:

  \begin{equation}
    cos\phi = \frac{ac_3 + bc_2}{(a^2 - b^2)sin\Theta} 
  \end{equation}

  If $cos\phi$ is positive, we choose solutions between $(-\pi/2, \pi/2)$. Else, we choose solutions in range $(-\pi, -\pi/2)\cup(\pi/2, \pi)$.  

  \section{Effect of MC sampling noise in simulated light curve}\label{apdx:mc-noise}
  Monte-Carlo Noise for a simulated light curve can be calculated by considering the flat part of the transit dip. The mean Monte-Carlo Noise of the light curves simulated in our training sample is $ \sigma_{MC} = 4.31\times10^{-3}\%$. The difference in the light curve as a result of change of geometry should ideally be greater than Monte-Carlo Noise. If the difference between any two light curves is less than that, then the effect of the shape deviation will not be captured in the simulated light curve.

  To assess this, we generate individual pairs from the set of simulated light curves, and evaluate the maximum difference in each pair ($max(|lc_1 - lc_2|)$). We find an insignificant fraction of pairs ($1.54\times10^{-5}$) that have residuals lower than $3\sigma_{MC}$. The mean difference between light curve pairs is $0.023$, which is far greater than the Monte-Carlo Noise. 
  Normalization of training light curve between 0 and 1 also scales the Monte-Carlo noise for some light curves, which correspond to shapes with small area. Even after scaling, there is a very small fraction of pairs ($1.4\times10^{-4}$) that have residuals lower than the scaled $3\sigma_{MC}$ ($0.0021$).  This analysis shows that Monte-Carlo noise is unlikely to have an impact in our training and predictions.

  \begin{figure}
    \includegraphics[width=1\linewidth]{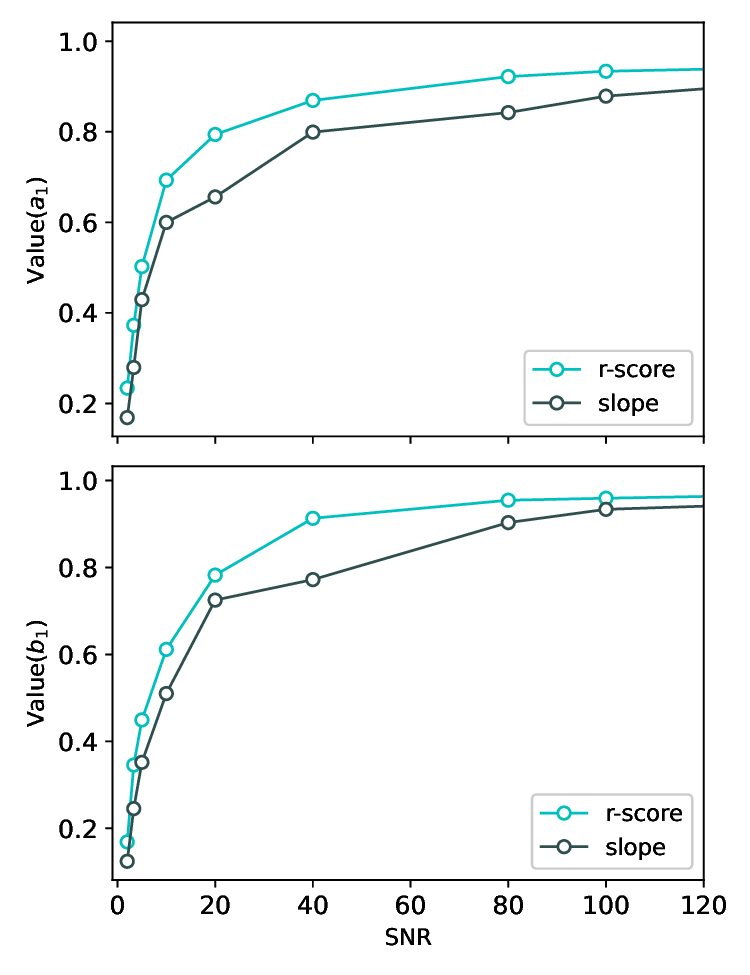}
    \caption{Impact of noisy light curves on prediction of parameter $a_1$ (top panel) and $b_1$ (bottom panel). The r-score and slope of the best fitted line between predicted and expected values of $a_1$ and $b_1$ are shown. These values are evaluated by training different SNR input light curves.
    \label{fig:noisy_lc}}
  \end{figure}

  \begin{figure}
        \centering
        \includegraphics[width=0.98\linewidth]{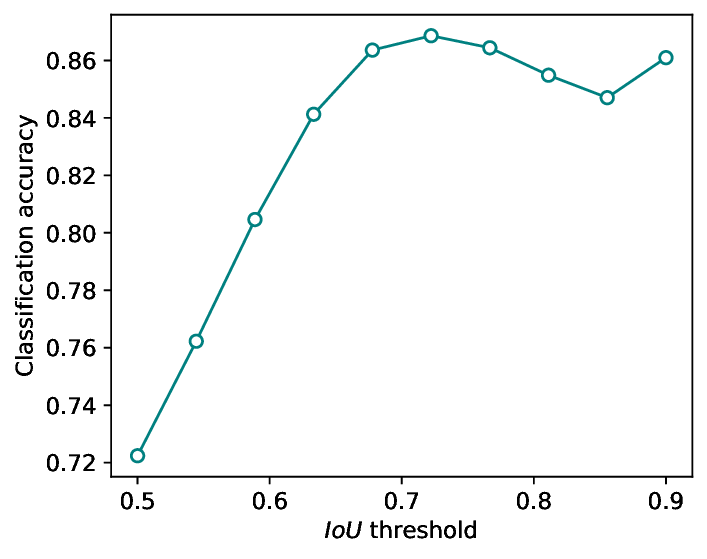}
        \caption{Accuracy of $IoU$ threshold classifier models for varying $IoU$ thresholds.}
        \label{fig:iou-threshold}
  \end{figure}

  \section{Impact of noise on learning of $a_1, b_1$}\label{apdx:lc-noise}
  This is a demonstration of how noise in the input light curve affects the training of Fourier parameters. We examine the performance of parameters $a_1$ and $b_1$. In the input light curves we add Gaussian noise such that SNR is constant for all training samples. Here SNR is defined as the transit depth divided by Gaussian $\sigma$. We observe that NN trained on noise-free light curves fail to predict parameters from noisy light curves. Therefore, the NN must be trained on noisy light curves.

  For a given SNR, we train the NN to learn the parameter $a_1, b_1$ from the noisy light curves. We calculate the r-score and slope from the predictions of the test sample. Fig. \ref{fig:noisy_lc} shows the variation of $m, r$ with SNR. For both $a_1$ and $b_1$, at high SNR, the $m, r$ values are close to the noise-free values. As we reduce the SNR, the performance degrades. The performance of the NN remains well upto an SNR $\sim 40$. Below SNR $\sim 20$, the performance degrades rapidly. This preliminary analysis shows that the scale and eccentricity of the overall effective ellipse can be estimated for many realistic transits, given that $35\%$ of Kepler Objects of Interest have SNR $>37$ \citep{kepleroi}.  A more detailed noise estimate will include dependence of noise on transit parameters, such as size and orbital period. The performance of various Fourier coefficients will also depend on the magnitude of deviation caused by individual Fourier coefficients to the transit light curve. We aim to explore these ideas in more detail in our upcoming works. 

  \section{IoU decision boundary cross-validation}\label{apdx:iou}
   We obtained the decision boundary in the $\mathcal{C}_{p}$-$\mathcal{C}_{s}$ plane using a support‑vector classifier. For the light curves in the training sample (24000 samples), we obtain the $\mathcal{C}_{p}$ from complexity network and complexity of reconstructed shape ($\mathcal{C}_{s}$). We assign the binary label (0 or 1) by applying a threshold to its Intersection‑over‑Union (IoU) value between the original shape and reconstructed shape. To assess the classifier we performed Monte‑Carlo cross‑validation (MCCV): in 40 iterations we randomly reshuffled the data, split it into 70\% training and 30\% validation sets, train the SVC on the training set, and then compute the classification accuracy on the validation set. Classification accuracy is then computed on the held-out test set (6000 samples). We repeat this procedure for IoU thresholds ranging from 0.5 to 0.9. Fig. \ref{fig:iou-threshold} shows that the highest validation accuracy is achieved at an $IoU$ threshold of 0.72.
    
\end{appendix}

\bibliography{compiled_bibliography}{}
\bibliographystyle{mnras}



\end{document}